\documentclass[lettersize,journal]{IEEEtran}
\usepackage{amsmath,amsfonts}
\usepackage{graphicx}
\usepackage{subcaption}

\usepackage{amsmath}
\usepackage{amssymb}
\usepackage{array}
\usepackage{subfig}
\usepackage{textcomp}
\usepackage{stfloats}
\usepackage{url}
\usepackage{verbatim}
\usepackage{cite}
\usepackage{enumitem}

\newtheorem{definition}{\textbf{Definition}}
\usepackage[ruled,vlined]{algorithm2e}
\usepackage[table,xcdraw]{xcolor}
\usepackage{multirow}

\begin{document}

\title{Evaluating of Machine Unlearning: Robustness Verification Without Prior Modifications} 

\author{Heng Xu, Tianqing Zhu*, Wanlei Zhou
    \thanks{
    *Tianqing Zhu is the corresponding author. Heng Xu is with the Centre for Cyber Security and Privacy and the School of Computer Science, University of Technology Sydney, Ultimo, NSW 2007, Australia (e-mail: heng.xu-2@student.uts.edu.au). Tianqing Zhu and Wanlei Zhou are with the City University of Macau, Macau (e-mail: tqzhu@cityu.edu.mo; wlzhou@cityu.edu.mo).
    }
}

\markboth{Journal of \LaTeX\ Class Files,~Vol.~14, No.~8, August~2021}%
{Shell \MakeLowercase{\textit{et al.}}: A Sample Article Using IEEEtran.cls for IEEE Journals}


\maketitle

\begin{abstract}
Machine unlearning, a process enabling pre-trained models to remove the influence of specific training samples, has attracted significant attention in recent years. While extensive research has focused on developing efficient unlearning strategies, the critical aspect of unlearning verification has been largely overlooked. Existing verification methods mainly rely on machine learning attack techniques, such as membership inference attacks~(MIAs) or backdoor attacks. However, these methods, not being formally designed for verification purposes, exhibit limitations in robustness and only support a small, predefined subset of samples. Moreover, dependence on prepared sample-level modifications of MIAs or backdoor attacks restricts their applicability in Machine Learning as a Service (MLaaS) environments. To address these limitations, we propose a novel robustness verification scheme without any prior modifications, and can support verification on a much larger set. Our scheme employs an optimization-based method to recover the actual training samples from the model. By comparative analysis of recovered samples extracted pre- and post-unlearning, MLaaS users can verify the unlearning process. This verification scheme, operating exclusively through model parameters, avoids the need for any sample-level modifications prior to model training while supporting verification on a much larger set and maintaining robustness. The effectiveness of our proposed approach is demonstrated through theoretical analysis and experiments involving diverse models on various datasets in different scenarios.
\end{abstract}


\section{Introduction}

Machine unlearning has recently attracted significant attention~\cite{DBLP:journals/csur/XuZZZY24,DBLP:conf/ndss/dayong2025,DBLP:conf/ccs/Chen000H022,DBLP:conf/sp/BourtouleCCJTZL21}, aiming to selectively remove the influence of specific training samples from trained models. Its growing popularity can be attributed to several factors, including the global trend toward stringent data protection regulations. For example, various legislators have enacted legislation granting individuals the ``right to be forgotten," which requires organizations to unlearn user data upon receiving requests~\cite{webpage:GDPR}.
Prior research on machine unlearning has primarily focused on developing efficient methods for removing specific samples from models. 
Despite the success achieved in machine unlearning strategies, The verification of unlearning emerged as a new requirement. Verification means a model provider can prove that the requested sample has been removed from the model, or user can test that his/her samples have been unlearned from the model successfully.  However, currently, users have limited ability to monitor the unlearning process and confirm if their data has been truly unlearned from the model~\cite{DBLP:journals/csur/XuZZZY24}.

Here are some prior research results on the verification of machine unlearning. Weng et al.~\cite{DBLP:journals/tifs/WengYDHWW24} proposed using trusted hardware to enforce proof of unlearning, though their method relies on trusted execution environments within Machine Learning as a Service (MLaaS). Cao et al.~\cite{DBLP:conf/sp/CaoY15} introduced data pollution attacks to assess if the model's performance was recovered to its original state after the unlearning process. Similarly, Guo et al.~\cite{DBLP:journals/tifs/GuoZHWJ24} evaluated whether the model provider truly performs machine unlearning by analyzing the performance of backdoor attacks. Other verification schemes are based on the distribution of model parameters~\cite{DBLP:conf/infocom/LiuXYWL22},  membership inference attacks (MIAs)~\cite{DBLP:journals/tifs/LiuXMW22,DBLP:conf/nips/JiaLRYLLSL23,DBLP:conf/nips/KurmanjiTHT23,DBLP:conf/aaai/FosterSB24,DBLP:conf/cvpr/ChenGL0W23},  model inversion attacks~\cite{DBLP:conf/aaai/GravesNG21,DBLP:journals/tifs/ChundawatTMK23}, theoretical analysis~\cite{DBLP:conf/icml/GuoGHM20}, or model accuracy~\cite{DBLP:conf/eccv/GolatkarAS20,DBLP:conf/cvpr/GolatkarARPS21,DBLP:conf/icml/BrophyL21,DBLP:conf/www/Wang0XQ22}.

However, those unlearning verification schemes have several limitations. Schemes based on model accuracy cannot reliably determine if targeted samples have been truly unlearned, since unlearning partial samples may not significantly affect model performance for those targeted samples~\cite{DBLP:conf/eccv/GolatkarAS20,DBLP:conf/cvpr/GolatkarARPS21,DBLP:conf/icml/BrophyL21,DBLP:conf/www/Wang0XQ22}. Schemes based on model inversion attacks do not support sample-level verification but rather only enable verification at the class level, as they can only recover class-level representation~\cite{DBLP:conf/aaai/GravesNG21,DBLP:journals/tifs/ChundawatTMK23}. Verification schemes, such as those based on MIAs, distribution of model parameters or theoretical analysis, are often ineffective due to unstable performance~\cite{DBLP:conf/nips/JiaLRYLLSL23,DBLP:conf/nips/KurmanjiTHT23}. Beyond these drawbacks, there are several common limitations:

\begin{itemize}
    \item Existing unlearning verification schemes always involve modifying samples prior to model training. For example, backdoor-based schemes require altering training samples by adding backdoor triggers~\cite{DBLP:journals/tifs/LiZYJWX23,DBLP:journals/tifs/GuoZHWJ24}. These triggers are then used to verify the unlearning process. Similarly, schemes relying on data pollution use modified, poisoned samples to confirm unlearning. However, if these modifications lose their effect during the learning or unlearning process, those verification schemes become ineffective.
    
    \item Most existing unlearning verification methods only support a small, predefined subset of samples. For example, schemes relying on backdoor or data poisoning techniques necessitate the preparation of samples used for verification before model training process. Consequently, those approaches are restricted to verifying only those pre-prepared samples, failing to consider the verification of unlearning for samples that were not modified prior to the training process~\cite{DBLP:journals/tifs/LiZYJWX23,DBLP:journals/tifs/GuoZHWJ24}. This constraint significantly narrows the scope of unlearning verification, potentially leaving critical gaps in the verification process.
    
    \item Nearly all existing verification schemes lack \textit{robustness}, which includes both long-term effectiveness and adaptability to varied-term unlearning conditions. To illustrate the concept of long-term effects, consider verification methods such as those based on parameter distribution, MIAs, backdoor, or data poisoning. These methods may initially demonstrate effectiveness in verifying unlearning immediately after the process. However, their reliability can diminish after subsequent model modifications like fine-tuning or pruning~\cite{DBLP:journals/csur/XuZZZY24,DBLP:journals/tifs/GuoZHWJ24,DBLP:journals/tifs/LiZYJWX23}. Moreover, in varied-term scenarios, many verification methods are context-dependent, only confirming successful unlearning under specific, predefined conditions. This limited scope increases the risk of false positives - erroneously confirming successful unlearning when the influence of targeted samples may actually persist or even amplify post-unlearning~\cite{DBLP:journals/hengxutbd,DBLP:journals/tnnlstarun,DBLP:conf/cvpr/ChenGL0W23,ijcai2024p40}.~(see Section~\ref{sec:finetuningisnotasolution}).
\end{itemize}


In this paper, we propose UnlearnGuard, a novel robustness verification scheme for machine unlearning that operates without requiring any prior modifications to training samples. Our approach is based on the observation that neural networks tend to memorize actual training samples. Therefore, we aim to directly extract these training samples from the model for verification purposes. 

\begin{enumerate}
    \item To address the first limitation, we redefine model training as a maximum margin problem, leveraging insights from implicit bias~\cite{DBLP:conf/nips/JiT20,DBLP:conf/iclr/LyuL20} and data reconstruction~\cite{DBLP:conf/nips/HaimVYSI22,DBLP:conf/nips/BuzagloHYVONI23}, we introduce our primary recovery loss component derived from Karush-Kuhn-Tucker (KKT) point conditions. This loss allows us to recover actual training samples from model used for verification without modifying the training samples. 
    \item To cover as many samples as possible that can be verified for unlearning, we propose an additional loss that allows further recovered samples to exhibit greater similarity to their original counterparts in training data space. This is achieved by minimizing the negative absolute outputs and applying a projection that constrains those newly recovered samples to remain within a specified range of pre-recovered samples. By enhancing this similarity, more recovered samples become suitable for verification. 
    \item To tackle the limitation of lacking robustness, all proposed loss components only use model parameters for recovery, without incorporating any other information and condition for verification. We also provide theoretical proof based on implicit bias~\cite{DBLP:conf/nips/JiT20,DBLP:conf/iclr/LyuL20}, demonstrating how the differences between samples recovered after successful unlearning and those before unlearning, providing a theoretical foundation for robustness property.
\end{enumerate}

It's worth noting that, in our experiments, we find that some fine-tuning-based machine unlearning schemes does not completely remove the influence of targeted samples. In fact, those schemes appear to deepen the impacts of targeted samples in the model. This result contrasts with previous findings~\cite{DBLP:journals/hengxutbd,DBLP:journals/tnnlstarun,DBLP:conf/cvpr/ChenGL0W23,ijcai2024p40}. Comparing with previous verification methods, our proposed verification scheme can further enhance the reliability of machine unlearning.

In summary, we make the following contributions.
\begin{itemize}
    \item We take the first step in addressing machine unlearning verification problem without prior sample-level modifications, considering both the robustness of the verification scheme and its capacity for more sample verifications.
    \item We propose an optimization-based method for recovering actual training samples from models, which enables users to verify machine unlearning by comparing samples recovered before and after the unlearning process.
    \item We provide theoretical proofs and analyses of our scheme based on implicit bias to demonstrate its effectiveness. 
\end{itemize}

\section{Preliminary}
\subsection{Related Works}
\subsubsection{Machine Unlearning}
 In response to \emph{the right to be forgotten}, the machine learning community has proposed various unlearning schemes. In our earlier survey, we comprehensively reviewed recent works on machine unlearning~\cite{DBLP:journals/csur/XuZZZY24}. This survey extensively covered several key aspects, including: (I) the motivation behind machine unlearning; (II) the goals and desired outcomes of the unlearning process; (III) a new taxonomy for systematically categorizing existing unlearning schemes based on their rationale and strategies; and (IV) the characteristics and drawbacks of existing unlearning verification schemes. 

Existing machine unlearning schemes are usually based on the following two techniques: \emph{data reorganization} and \emph{model manipulation}~\cite{DBLP:journals/csur/XuZZZY24}. Data reorganization refers to restructuring the training dataset to facilitate efficient machine unlearning. Cao et al.~\cite{DBLP:conf/sp/CaoY15} transformed the training dataset into summations forms, allowing updates to summations rather than retraining the entire model. Bourtoule et al.~\cite{DBLP:conf/sp/BourtouleCCJTZL21} proposed one ``Sharded, Isolated, Sliced, and Aggregated" (SISA) framework, where data is partitioned into disjoint shards, and sub-models are retrained only on the shards containing the sample to be unlearned. Similar schemes are used in graph tasks~\cite{DBLP:conf/ccs/Chen000H022}. Model manipulation-based schemes usually adjust the model's parameters directly. Guo et al.~\cite{DBLP:conf/icml/GuoGHM20} proposed a certified removal scheme based on the influence function~\cite{DBLP:conf/icml/KohL17} and differential privacy~\cite{DBLP:journals/tkde/ZhuLZY17}. Schemes in~\cite{DBLP:journals/tifs/ZhangZZXZ23,DBLP:conf/www/LinGDNK024} consider unlearning requests in federated learning setting. Kurmanji et al.~\cite{DBLP:conf/nips/KurmanjiTHT23} and Chen et al.~\cite{DBLP:conf/nips/ChenYXBHHFZWL23} use machine unlearning techniques to address bias issues and resolve ambiguities or confusion in machine learning models. 
  
\subsubsection{Machine Unlearning Verification}
\label{sec:machineunlearningverification}
Current schemes for verifying machine unlearning can be broadly categorized into: \emph{empirical evaluation} and \emph{theoretical analysis}~\cite{DBLP:journals/csur/XuZZZY24}.

Empirical evaluation schemes mainly employ attack methods to evaluate how much information about samples targeted for unlearning remains within the model. For example, model inversion attacks have been used in studies, such as~\cite{DBLP:conf/aaai/GravesNG21,DBLP:journals/tifs/ChundawatTMK23}, while membership inference attacks~(MIAs) were employed in ~\cite{DBLP:conf/nips/JiaLRYLLSL23,DBLP:conf/nips/KurmanjiTHT23,DBLP:conf/aaai/FosterSB24,DBLP:conf/cvpr/ChenGL0W23}. Cao et al.~\cite{DBLP:conf/sp/CaoY15} introduced data pollution attacks to verify whether the model performance was restored to its initial state after the unlearning process. Similar strategies involving backdoor attacks were adopted in~\cite{DBLP:journals/popets/SommerSWM22,DBLP:journals/tifs/LiZYJWX23,DBLP:journals/tifs/GuoZHWJ24}. Beyond attack-based methods, Liu et al.~\cite{DBLP:conf/infocom/LiuXYWL22}, Wang et al.~\cite{DBLP:conf/www/Wang0XQ22} and Brophy et al.~\cite{DBLP:conf/icml/BrophyL21} used accuracy metrics, whereas Baumhauer et al.~\cite{DBLP:conf/sp/BourtouleCCJTZL21}, Golatkar et al.~\cite{DBLP:conf/cvpr/GolatkarAS20,DBLP:conf/cvpr/GolatkarARPS21}, and Liu et al.~\cite{DBLP:conf/iwqos/LiuMYWL21} measured the similarity of distributions of pre-softmax outputs. Additionally, some efforts have been made to compare the similarity between the parameter distributions of a model after unlearning and a model after retraining from scratch~\cite{DBLP:conf/infocom/LiuXYWL22}. On the other hand, theoretical analysis schemes typically focus on ensuring that the unlearning operation effectively removes the targeted sample information from model~\cite{DBLP:conf/icml/GuoGHM20}.


\subsubsection{Discussion of Related Works} Despite the progress made in machine unlearning, current verification methods still face significant challenges. The limitations of existing verification schemes can be summarized as follows: 

\begin{itemize}
    \item Accuracy-Based Verification: Schemes relying on accuracy metrics cannot adequately determine whether samples have been unlearned, since the accuracy of unlearned samples does not necessarily reflect the true effectiveness of the unlearning process~\cite{DBLP:conf/eccv/GolatkarAS20,DBLP:conf/cvpr/GolatkarARPS21,DBLP:conf/icml/BrophyL21,DBLP:conf/infocom/LiuXYWL22,DBLP:conf/www/Wang0XQ22}.
    \item Similarity-Based Verification: Approaches that measure the similarity of distributions, such as pre-softmax outputs or parameters, have been shown to be ineffective in~\cite{DBLP:conf/uss/ThudiJSP22}.
    \item Attack-Based Verification: Schemes based on model poisoning attacks, such as data pollution~\cite{DBLP:conf/sp/CaoY15} and backdoor attacks~\cite{DBLP:journals/popets/SommerSWM22,DBLP:journals/tifs/LiZYJWX23,DBLP:journals/tifs/GuoZHWJ24} require modification of a subset of samples before model training. Consequently, those methods can only verify the unlearning process for specific samples that were poisoned during training, but not for any unspecified data after training.
    \item Theoretical Analysis Verification: While theoretical analysis schemes~\cite{DBLP:conf/icml/GuoGHM20} can ensure the unlearning results, they are often constrained by specific unlearning strategies and are less effective with complex models and large datasets.
\end{itemize}

Most importantly, the majority of existing verification methods, including data pollution~\cite{DBLP:conf/sp/CaoY15} and backdoor attack-based~\cite{DBLP:journals/popets/SommerSWM22,DBLP:journals/tifs/GuoZHWJ24}, fail to consider long-term verification. While these methods may be effective immediately after unlearning, they become invalid once the model undergoes a fine-tuning or pruning process. We will discuss this in detail in Section~\ref{sec:existenceschemes}.

\subsection{Background}

Machine Learning as a Service (MLaaS) primarily involves two key entities: the \textit{data provider} and the \textit{model provider}. The data provider submits their data to the model provider, who then uses this data for model training. We denote the dataset of the data provider as $\mathcal{D}=\left\{\left(\mathbf{x}_{1}, y_{1}\right), \left(\mathbf{x}_{2}, y_{2}\right),...,\left(\mathbf{x}_{n}, y_{n}\right) \right\} \subseteq \mathbb{R}^{d} \times \mathbb{R}$, where each sample $\mathbf{x}_{i} \in \mathcal{X}$ is a $d$-dimensional vector, $y_{i} \in \mathcal{Y}$ is the corresponding label, and $n$ is the size of $\mathcal{D}$. Let $\mathcal{A}$ be a (randomized) learning algorithm that trains on $\mathcal{D}$ and outputs a model $M$. The model $M$ is given by $M = \mathcal{A}(\mathcal{D})$, where $M \in \mathcal{H}$ and $\mathcal{H}$ is the hypothesis space. After the training process, data providers may wish to unlearn specific samples from the trained model and submit an unlearning request. Let $\mathcal{D}_{u} \subset \mathcal{D}$ represent the subset of the training dataset that the data provider wishes to unlearn. The complement of this subset, $\mathcal{D}_{r}=\mathcal{D}_{u}^{\complement} = \mathcal{D} \setminus \mathcal{D}_{u}$, represents the data that the provider wishes to retain.

\begin{definition}[Machine Unlearning~\cite{DBLP:journals/csur/XuZZZY24}]
    \label{Definition:Machineunlearning}
    Consider a set of samples that a data provider wishes to unlearn from an already-trained model, denoted as $\mathcal{D}_{u}$. The unlearning process, represented as $\mathcal{U}(M, \mathcal{D}, \mathcal{D}_{u})$, is a function that takes an already-trained model $M = \mathcal{A}(\mathcal{D})$, the training dataset $\mathcal{D}$, and the unlearning dataset $\mathcal{D}_{u}$, and outputs a new model $M_u$. This process ensures that the resulting model, $M_u$, behaves as if it had never been trained on $\mathcal{D}_{u}$.
\end{definition}

After the machine unlearning process, a verification procedure $\mathcal{V}(\cdot)$ is employed to determine whether the requested samples have been successfully unlearned from the model. Typically, data providers lack the capability to perform this verification independently. For example, verification schemes based on MIAs often require the training of attack models using multiple shadow models, which can be too resource-intensive for data providers. Consequently, verification operations in MLaaS are usually conducted by a trusted third party. With the assistance of this trusted third party, a distinguishable check $\mathcal{V}(\cdot)$ will be conducted to ensure that $\mathcal{V}\left(M, \mathcal{D}_u\right) \neq \mathcal{V}\left(M_u,\mathcal{D}_u\right)$.

\section{Problem Definition}

\subsection{Existing Verification Schemes}
\label{sec:existenceschemes}
\begin{figure}
    \centering
    \includegraphics[width=1\linewidth]{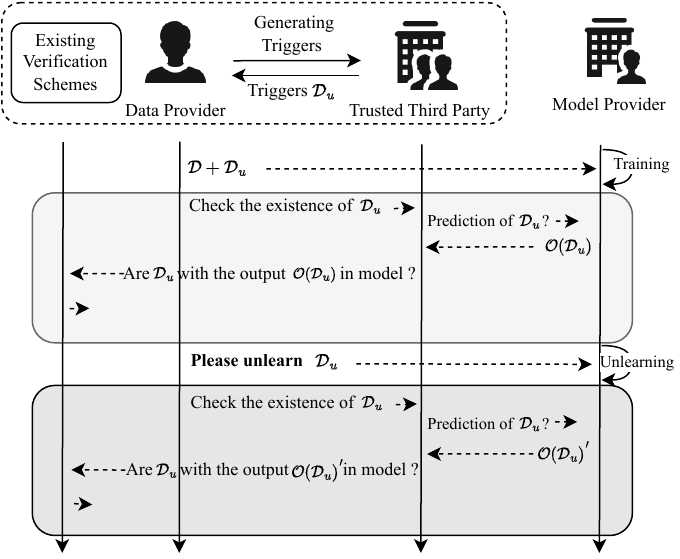}
    \caption{Existing verification scheme process.}
    \label{fig:existingverificationscheme}
\end{figure}

Designing an effective and efficient verification scheme for machine unlearning is difficult. As discussed in Section~\ref{sec:machineunlearningverification}, verification schemes based on accuracy, theoretical analysis and distribution similarity have consistently proven ineffective in~\cite{DBLP:journals/csur/XuZZZY24,DBLP:conf/uss/ThudiJSP22}. In this Section, we further highlight the critical limitations of attack-based verification schemes to illustrate the novelty of our scheme. Figure~\ref{fig:existingverificationscheme} illustrates the main ideas of commonly used attack-based verification schemes.

As shown in Figure~\ref{fig:existingverificationscheme}, attack-based unlearning verification schemes typically involve three roles, including data provider, model provider and one trusted third party. Data provider first pre-selects a subset of triggers $\mathcal{D}_u$ before model training. For example, in schemes based on data pollution and backdoor attacks, triggers are often generated by a trusted third party~\cite{DBLP:conf/sp/CaoY15,DBLP:journals/popets/SommerSWM22,DBLP:journals/tifs/GuoZHWJ24}. In MIAs-based verification schemes, data providers directly select partial training samples as those triggers~\cite{DBLP:conf/nips/JiaLRYLLSL23,DBLP:conf/nips/KurmanjiTHT23,DBLP:conf/aaai/FosterSB24,DBLP:conf/cvpr/ChenGL0W23}. Based on those triggers, the verification process can be summary as the following steps:

\begin{itemize}
    \item \textbf{Trigger Integration}: Triggers $\mathcal{D}_u$, either generated from a trusted third party or selected directly from the data provider’s own dataset, are added to the training dataset. The model is then trained on this combined dataset.
    \item \textbf{Initial Prediction}: The data provider sends the verification request to the trusted third party, and the trusted third party queries the model's prediction for these triggers. The model provider returns the predictions $O(\mathcal{D}_u)$. Existing attack-based verification scheme outputs the verification result $\mathcal{V}(O(\mathcal{D}_u))$ for $\mathcal{D}_u$ based on $O(\mathcal{D}_u)$.
    \item \textbf{Unlearning Request}: The data provider submits an unlearning request for those triggers $\mathcal{D}_u$.
    \item \textbf{Post-Unlearning Prediction}: The data provider and the trusted third party repeat the steps in the initial prediction phase and output the verification result $\mathcal{V}(O(\mathcal{D}_u)')$.
    \item \textbf{Verification}: By comparing the returned predictions before and after unlearning, $\mathcal{V}(O(\mathcal{D}_u))$ and $\mathcal{V}(O(\mathcal{D}_u)')$ respectively, the data provider determines if the model has undergone the unlearning process.
\end{itemize}

However, those attack-based machine unlearning verification schemes mainly have the following limitations:
\begin{itemize}
    \item \textbf{Sample-Level Modification before Model Training}: Ensuring effective verification often requires incorporating a large number of modified samples, such as backdoored samples, which increases computational costs.
    \item \textbf{Support Only a Small, Predefined Subset of Samples}: Some schemes use pre-embedded patterns for verification, like backdoor triggers, limiting the process to only those samples. This means the number of verifications must be considered before training, and once these samples are used up, no further verification is possible.
    \item \textbf{Lacking Robustness Verification}: These schemes focus solely on verifying the unlearning process immediately and lack robustness, meaning the model cannot be fine-tuned or pruned for new demands after unlearning.
    \item \textbf{Model Performance Degradation and Security Risks}: Verification schemes based on data pollution and backdoor attacks rely on poisoned samples, inheriting the drawbacks of poisoning methods. This can harm model performance and introduce security risks, limiting their adoption for verification.
\end{itemize}


\subsection{Threat Model and Goals}

In MLaaS, there are two main roles: data providers and model providers. Data providers also act as verifiers of unlearning, while model providers are responsible for executing it. The data provider shares its dataset with the model provider, who trains a model using a learning algorithm. After training, beyond making regular predictions, the data provider can submit unlearning requests to remove their data from the model. However, model providers may not always be fully trustworthy in performing unlearning, either because this process is time-consuming, or large-scale unlearning requests could negatively impact model performance~\cite{DBLP:journals/tifs/GuoZHWJ24,DBLP:journals/popets/SommerSWM22}. The background and goals of data providers are as follows:

\begin{itemize}
    \item \textbf{Background:} Data providers only have the ability to upload their training data or send prediction, unlearning requests to the model providers.
    \item \textbf{Goal:} After submitting the unlearning request, data providers want to confirm whether their data has been truly unlearned from the model.
\end{itemize}

Meanwhile, the background and goals of model providers are as follows:

\begin{itemize}
    \item \textbf{Background:} Model provider can collect training data from the data provider and train the model.
    \item \textbf{Goal:} After receiving the unlearning requests from data provider, model providers prefer to avoid executing unlearning as much as possible to protect their own interests. 
\end{itemize}


This paper proposes a method for data providers to verify whether their samples have been successfully unlearned from the trained model. Specifically, we have four aims related to machine unlearning verification.
\begin{itemize}
    \item \textbf{No Pre-defined Modifications}: Develop a verification scheme that enables data providers to confirm the execution of the unlearning process without depending on any pre-defined sample-level modifications.
    \item \textbf{More Sample Coverage}: Design a scheme supporting unlearning verification for nearly all samples involved in the training process.
    \item \textbf{Robustness Verification}: Address the need for robustness by supporting immediate post-unlearning verification and enabling verification in scenarios where the model undergoes further changes after unlearning~(e.g., further fine-tuning or model pruning).
    \item \textbf{Preserving Model Usability}: Ensure that the verification scheme does not negatively impact: model performance, security, and training efficiency.
\end{itemize}

We assume that the data provider conducts verification with the help of a trusted third party. This assumption reflects real-world MLaaS scenarios where data providers often lack the capability to independently verify the unlearning processes~\cite{DBLP:journals/tifs/GuoZHWJ24}. The trusted third party is granted access to the trained model for verification purposes upon receiving unlearning requests. Additionally, we assume that the model provider may carry out further model modifications~(e.g., fine-tuning and pruning) after executing the unlearning process.

\begin{figure}[!t]
    \centering
    \includegraphics[width=1\linewidth]{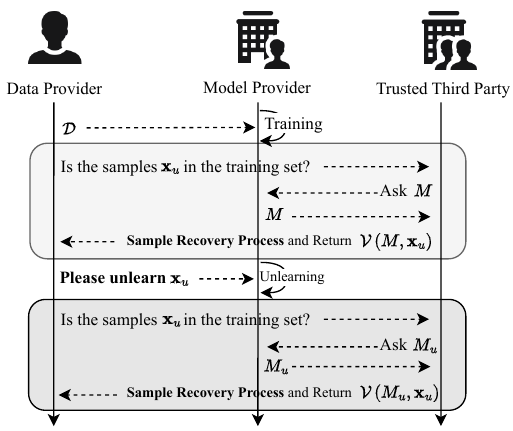}
    \caption{Our verification scheme process.}
    \label{fig:ourscheme}
\end{figure}

\section{Methodology}
\subsection{Overview}

Current unlearning verification schemes typically depend on additional information to verify the unlearning process, such as pre-embedded backdoors~\cite{DBLP:journals/tifs/GuoZHWJ24,DBLP:journals/popets/SommerSWM22}. However, these schemes become ineffective if this additional information is disrupted during subsequent model fine-tuning or pruning. Therefore, it is crucial to verify the unlearning process only using the model itself and investigate if the information can still be recovered from the model parameters post-unlearning.

In this paper, we propose a robustness verification scheme without prior modifications, named UnlearnGuard, which verifies machine unlearning based only on the model and possesses robustness properties. As illustrated in Figure~\ref{fig:ourscheme}, UnlearnGuard directly verifies the unlearning process by examining whether the model parameters contain information about the unlearning samples. Before and after unlearning, the data provider sends a verification request about $\mathbf{x}_u$ to trusted third party. The trusted third party ask model from model provider and attempts to recover $\mathbf{x}_u$ from it. Based on two recovery results, the data provider can determine if the data was truly unlearned. Our approach distinguishes UnlearnGuard from the scheme illustrated in Figure~\ref{fig:existingverificationscheme} as UnlearnGuard relies only on the model itself rather than any prior sample-level modifications.

We describe our scheme based on the following two subsections: \emph{unlearning verification process} and \emph{sample recovery process}. In Section~\ref{sec:unlearningverificationprocess},  we introduce the main workflow of our entire unlearning verification process, which includes the pre-unlearning process and the post-unlearning process, encompassing all steps shown in Figure~\ref{fig:ourscheme}. Next, in Section~\ref{sec:datareconstruction}, we explain how to recover actual unlearning samples from the model, which is the most important part of our scheme and supports the workflow of Section~\ref{sec:unlearningverificationprocess}.

\subsection{Unlearning Verification Process}
\label{sec:unlearningverificationprocess}

In this section, we will describe the whole workflow of our unlearning verification process, including three steps. 

\subsubsection{Model Training and Pre-verification} Data providers can submit their datasets to MLaaS for model training, enhancing accessibility and efficiency in machine learning deployment. After training, with the help of a trusted third party, data providers can perform pre-verification to confirm the presence of their data within the trained model. While this verification step enhances the comprehensiveness of our proposed scheme, it is not mandatory for practical implementations.

To conduct this, data providers send sample $\mathbf{x_u}$  to a trusted third party, requesting confirmation of $\mathbf{x_u}$'s presence in the model. The trusted third party then asks the MLaaS model provider for the trained model $M$ and attempts to recover the samples using the scheme described in Section~\ref{sec:datareconstruction}. This verification process outputs a result $\mathcal{V}\left(M,\mathbf{x_u}\right)$.

\subsubsection{Unlearning Request and Execution} To initiate unlearning, the data provider sends a request regarding samples $\mathbf{x}_{u}$. The model provider locates and removes sample $\mathbf{x}_{u}$ from the dataset and executes the unlearning process.

\subsubsection{Re-query and Verification} 
Following the unlearning request, the data provider again inquires about $\mathbf{x_u}$'s presence in the model. The trusted third party returns the output $\mathcal{V}\left(M_{u},\mathbf{x_u}\right)$, where $M_{u}$ represents the updated model after unlearning. The data provider then compares $\mathcal{V}\left(M,\mathbf{x_u}\right)$ and $\mathcal{V}\left(M_{u},\mathbf{x_u}\right)$ to determine whether the model provider has successfully executed the unlearning operation. In Algorithm~\ref{algorithm:verification}, we provide a detailed process for this process. 

\begin{algorithm}[!t]
	\small
 	\caption{Unlearning Verification Process}
	\label{algorithm:verification}
	\LinesNumbered 
	\KwIn{Model $M$ and sample $\mathbf{x}_u$ that need to be unlearned}
	\KwOut{Whether model provider has unlearned samples $\mathbf{x}_u$}
        \textbf{Data provider executes}: \\
        ~~~~~Query if $\mathbf{x}_u$ is in the model.\\
        ~~~~~$\mathcal{V}\left(M,\mathbf{x_u}\right)$ $\leftarrow$ receive results from trust third party \\
        ~~~~~Sending the unlearning request regarding samples $\mathcal{D}_u$.\\
        ~~~~~Re-query if $\mathbf{x}_u$ is in the model.\\
        ~~~~~$\mathcal{V}\left(M_{u},\mathbf{x_u}\right)$ $\leftarrow$ receive results from trust third party \\
        ~~~~~\If{$\mathcal{V}\left(M,\mathbf{x_u}\right)$ == True and $\mathcal{V}\left(M_{u},\mathbf{x_u}\right)$  == False}{
              ~~~~~\Return {Unlearning process has executed.}\\
             }
        ~~~~~\If{$\mathcal{V}\left(M,\mathbf{x_u}\right)$ == True and $\mathcal{V}\left(M_{u},\mathbf{x_u}\right)$  == True}{
             ~~~~~\Return {Unlearning process has not executed.}\\
        }   
        \textbf{Trusted third party executes}: \\
	~~~~~\textbf{Upon} receiving a verification request from data provider{\\
             ~~~~~~~~~~Send a request to model provider to query the model.\\
             ~~~~~~~~~~Try to recovering sample $\mathbf{x}_u$ based on Algorithm~\ref{algorithm:recover}.\\
             ~~~~~~~~~~\Return {recover results $\mathcal{V}()$}.\\
        }
        \textbf{Model provider executes}: \\
	~~~~~\textbf{Upon} receiving a request to query the model:{\\
             ~~~~~~~~~~\Return {$M$}.\\
        }
\end{algorithm}

In Algorithm~\ref{algorithm:verification}, lines 2-3, denote the pre-verification process. Specially, the initial result $\mathcal{V}\left(M,\mathbf{x_u}\right)$ should be True. Line 4 shows the data provider sending an unlearning request, followed by another query for the result related to $\mathbf{x}_u$ (lines 5-6). If $\mathcal{V}\left(M_{u},\mathbf{x_u}\right)$ is False, it confirms that the model provider has truly executed the unlearning operation~(lines 7-8); otherwise, it suggests that the model provider has not executed the unlearning operation~(lines 9-10). In our scheme, we define the function $\mathcal{V}$ as:

\begin{equation}
    \mathcal{V}(\mathbf{x}_u,\left\{\mathbf{x}_i\right\}_{i=1}^m) = \bigvee_{i=1}^m \left( \text{SSIM}(\mathbf{x}_u, \mathbf{x}_i) \gg \eta \right)
\end{equation}
where $\vee$ denotes the logical OR operation and $m$ is the number of recovered samples.

\subsection{Sample Recovery Process}
\label{sec:datareconstruction}
In the previous section, we explained the process of verifying unlearning, mainly based on comparing the recovery results before and after unlearning. This section describes how to recover actual training samples encoded within a trained model, leveraging implicit bias and date reconstruction. 

We begin by studying simple models before advancing to the analysis of more complex deep models in Section~\ref{sec:theoreticalanalysis}. Let $D=\left\{\left(\mathbf{x}_i, y_i\right)\right\}_{i=1}^n \subseteq \mathbb{R}^d \times\{-1,1\}$ be a binary classification training dataset.  Consider a neural network $M(\boldsymbol{\theta} ; \cdot)$: $\mathbb{R}^d \rightarrow \mathbb{R}$ parameterized by $\boldsymbol{\theta} \in \mathbb{R}^p$. For a given loss function $\ell: \mathbb{R} \rightarrow \mathbb{R}$, the empirical loss of $M(\boldsymbol{\theta} ; \cdot)$ on the dataset $D$ is given by: $\mathcal{L}(\boldsymbol{\theta}):=\sum_{i=1}^n \ell\left(y_i M\left(\boldsymbol{\theta} ; \mathbf{x}_i\right)\right)$. Let us consider the logistic loss, also known as binary cross-entropy, which is defined as $\ell(q)=\log \left(1+e^{-q}\right)$.

Directly recovering samples from the above-defined model poses significant challenges. To address this, we reformulate model training problem into a maximum margin problem based on the implicit bias theory discussed by Ji et al.~\cite{DBLP:conf/nips/JiT20} and Lyu et al.~\cite{DBLP:conf/iclr/LyuL20}, which simplifies the process of recovering samples from the model.

\textbf{Theorem 1}~(Paraphrased from Ji et al.~\cite{DBLP:conf/nips/JiT20}, Lyu et al.~\cite{DBLP:conf/iclr/LyuL20}): Consider a homogeneous neural network $M(\boldsymbol{\theta} ; \cdot)$. When minimizing the logistic loss over a binary classification dataset $\left\{\left(\mathbf{x}_i, y_i\right)\right\}_{i=1}^n$ using gradient flow, and assuming there exists a time $t_0$ such that $\mathcal{L}\left(\boldsymbol{\theta}\left(t_0\right)\right)<1$, then, gradient flow converges in direction to a first-order stationary point of the following maximum-margin problem:
\begin{equation}
\label{equ:kkt}
\min _{\boldsymbol{\theta}^{\prime}} \frac{1}{2}\left\|\boldsymbol{\theta}^{\prime}\right\|^2 \quad \text { s.t. } \quad \forall i \in[n] \quad y_i M\left(\boldsymbol{\theta}^{\prime} ; \mathbf{x}_i\right) \geq 1
\end{equation}

In this theorem, homogeneous networks are defined with respect to their parameters $\boldsymbol{\theta}$. Specifically, a network $M$ is considered homogeneous if there exists $L>0$ such that for any $\alpha>0$, $\boldsymbol{\theta}$ and $\mathbf{x}$, the relationship $M(\alpha \boldsymbol{\theta} ; \mathbf{x})=\alpha^L M(\boldsymbol{\theta} ; \mathbf{x})$ holds.  This means that scaling the parameters by any factor $\alpha>0$ results in scaling the outputs by $\alpha^L$. Gradient flow is said to converge in the direction to  $\tilde{\boldsymbol{\theta}}$ if $\lim _{t \rightarrow \infty} \frac{\boldsymbol{\theta}(t)}{\|\boldsymbol{\theta}(t)\|}=\frac{\tilde{\boldsymbol{\theta}}}{\|\boldsymbol{\boldsymbol { \theta }}\|}$, where $\boldsymbol{\theta}(t)$ is the parameter vector at time $t$. $\mathcal{L}\left(\theta\left(t_0\right)\right)<1$ means that there exists time $t_0$ at which the network classifies all the samples correctly.

This theorem describes how optimization algorithms, such as gradient descent, tend to converge to specific solutions that can be formalized as Karush-Kuhn-Tucker (KKT) conditions, enabling data reconstruction from the model using those conditions~\cite{DBLP:conf/nips/HaimVYSI22}. The reconstruction loss can be defined as following. Derivation details can be found in Section~\ref{sec:theoreticalanalysis}.

\begin{equation}
    \label{equ:others}
    \begin{aligned}
    L_{\text {reconstruct }}&\left(\left\{\mathbf{x}_i\right\}_{i=1}^m,\left\{\lambda_i\right\}_{i=1}^m\right)\\
    &=\alpha_1 L_{\text {stationary }}+\alpha_2 L_\lambda+\alpha_3 L_{\text {additional}}\\
    \text{s.t.}&~~~ L_{\text {stationary }}=\left\|\boldsymbol{\theta}-\sum_{i=1}^m \lambda_i y_i \nabla_{\boldsymbol{\theta}} M\left(\boldsymbol{\theta} ; \mathbf{x}_i\right)\right\|_2^2 \\
    &~~~~~ L_\lambda=\sum_{i=1}^m \max \left\{-\lambda_i, 0\right\}
    \end{aligned}
\end{equation}

where $m$ denotes the cardinality of the sample set to be reconstructed, typically set to $m \geq 2 n$ in their experimental setting. The loss $L_{\text {stationary }}$ represents the stationarity condition satisfied by the parameters at the KKT point, while $L_\lambda$ represents the dual feasibility condition. $L_{\text {additional}}$ denotes some supplementary constraints predicated on image attributes, such as ensuring pixel values remain between $[0, 1]$. 

Using the above $L_{\text {reconstruct }}$ can recover the actual sample from the model, which is our main purpose. However, only using this loss to reconstruct samples exhibits limitations in the context of partial verification. Specifically, it is constrained to reconstructing only those training samples that almost lie on the decision boundary margin. Consequently, when employed to verify the unlearning process, its efficacy is limited to a subset of samples—those situated on the margin. To expand the scope of reconstruction, we introduce a novel loss term, denoted as the prior information loss, $L_{\text{prior}}$.

The introduction of this new loss term aims to incorporate classification information pertaining to the samples targeted for recovery. Let $\left\{\mathbf{x}_i\right\}_{i=1}^m$ represent the samples recovered based on the aforementioned scheme. Our objective is to ensure that the subsequently recovered samples $\left\{\hat{\mathbf{x}_i}\right\}_{i=1}^m$ exhibit greater similarity to their counterparts in the training data space:

\begin{equation}
    \text{maximize} \left( M_{y_i}(\hat{\mathbf{x}_i}; \boldsymbol{\theta}) \right) = - \left| M_{y_i}(\hat{\mathbf{x}_i}; \boldsymbol{\theta}) \right|
\end{equation}

\begin{algorithm}[!t]
	\small
	\caption{Sample Recovery Process}
    \label{algorithm:recover}
	\LinesNumbered 
	\KwIn{The trained model $M$, iteration number $T_{1}$ and $T_{2}$.}
	\KwOut{Recovered Samples $\mathbf{x}^{'}$}
        ~~~~~\For{$t=0; t<T_{1}; t++$}{
            ~~~~~$L_{\text {reconstruct}}=\alpha_1 L_{\text {stationary }}+\alpha_2 L_\lambda$\\
            ~~~~~Optimizing $\left(\left\{\mathbf{x}_i\right\}_{i=1}^m,\left\{\lambda_i\right\}_{i=1}^m\right)$ based on $L_{\text {reconstruct}}$.\\

        }
        ~~~~~$\left\{\hat{\mathbf{x}_i}\right\}_{i=1}^m$ = $\left\{\mathbf{x}_i\right\}_{i=1}^m$\\
        ~~~~~\For{$t=0; t<T_{2}; t++$}{
            ~~~~~$L_{\text {reconstruct}}=\alpha_1 L_{\text {stationary }}+\alpha_2 L_\lambda+\alpha_3 L_{\text {prior}}$ \\
            ~~~~~Optimizing $\left(\left\{\hat{\mathbf{x}_i}\right\}_{i=1}^m,\left\{\lambda_i\right\}_{i=1}^m\right)$ based on $L_{\text {reconstruct}}$.\\
            ~~~~~Projecting within $[ \left\{\mathbf{x}_i\right\}_{i=1}^m - \epsilon, \left\{\mathbf{x}_i\right\}_{i=1}^m + \epsilon ]$
        }
    \Return {$\mathbf{x}^{'} = \hat{\mathbf{x}}$}\\
\end{algorithm}

Specifically, we aim to maximize $M_{y_i}(\hat{\mathbf{x}_i}; \boldsymbol{\theta})$ assigned to the predicted class for each sample $\hat{\mathbf{x}_i}$ by minimizing its negative absolute value. It is noteworthy that this optimization process does not utilize the original labels. Instead, it optimizes the model's output confidence (logits) for samples $\left\{\hat{\mathbf{x}_i}\right\}_{i=1}^m$, aligning with our hypothesis of implementing unlearning verification using only the model parameters.

Finally, we define our recovery loss as:

\begin{equation}
    \label{equ:ours}
    \begin{aligned}
    &L_{\text {reconstruct}}=\alpha_1 L_{\text {stationary }}+\alpha_2 L_\lambda+\alpha_3 L_{\text {prior}}\\
    \text{s.t.}&~~~L_{\text {stationary }}=\left\|\boldsymbol{\theta}-\sum_{i=1}^m \lambda_i y_i \nabla_{\boldsymbol{\theta}} M\left(\boldsymbol{\theta} ; \hat{\mathbf{x}_i}\right)\right\|_2^2 \\
    &~~~ L_\lambda=\sum_{i=1}^m \max \left\{-\lambda_i, 0\right\}~~
    L_{\text {prior}} =- \left| M_{y_i}(\hat{\mathbf{x}_i}; \boldsymbol{\theta}) \right|\\
    \end{aligned}
\end{equation}

where we eliminate  the $L_{\text {additional}}$ loss term and incorporate $ L_{\text {prior}}$ to enhance the fidelity of recovered samples.

To mitigate excessive deviation of the newly recovered samples $\left\{\hat{\mathbf{x}_i}\right\}_{i=1}^m$ from $\left\{\mathbf{x}_i\right\}_{i=1}^m$ in the data space, we introduce a projection function, \texttt{project\_to\_bounds}, applied after each optimization epoch. This function constrains the pixel values of $\left\{\hat{\mathbf{x}_i}\right\}_{i=1}^m$ to within the range $[ \left\{\mathbf{x}_i\right\}_{i=1}^m - \epsilon, \left\{\mathbf{x}_i\right\}_{i=1}^m + \epsilon ]$. Our recovery algorithm is shown in Algorithm~\ref{algorithm:recover}. 

In lines 1-3, we initially optimize $\left\{\mathbf{x}_i\right\}_{i=1}^m$ and $\left\{\lambda_i\right\}_{i=1}^m$ utilizing the loss $L_{\text {reconstruct}}=\alpha_1 L_{\text {stationary }}+\alpha_2 L_\lambda$. This preliminary phase ensures that $\left\{\mathbf{x}_i\right\}_{i=1}^m$ achieves a basic level of recovery. Subsequently, we proceed to optimize $\left\{\mathbf{x}_i\right\}_{i=1}^m$ further to recover samples $\left\{\hat{\mathbf{x}_i}\right\}_{i=1}^m$ using the augmented loss function~(lines 5-8). At each optimization epoch, we project the recovered samples onto a constrained space~(line 8). This ensures they remain within a specified range of the pre-recovered samples while capturing more essential features required for effective unlearning verification.

\subsection{Theoretical Analysis}
\label{sec:theoreticalanalysis}

Our machine unlearning verification scheme builds upon existing works~\cite{DBLP:conf/iclr/LyuL20,DBLP:conf/nips/JiT20,DBLP:conf/nips/HaimVYSI22,DBLP:conf/nips/BuzagloHYVONI23}, extending their insights to the context of machine unlearning. We provide a rigorous theoretical basis for our verification method, incorporating concepts from optimization theory and functional analysis.

Let $D=\left\{\left(\mathbf{x}_i, y_i\right)\right\}_{i=1}^n \subseteq \mathbb{R}^d \times\{-1,1\}$ be a binary classification training dataset.  Consider a neural network $M(\boldsymbol{\theta} ; \cdot)$: $\mathbb{R}^d \rightarrow \mathbb{R}$ parameterized by $\theta \in \mathbb{R}^p$. The training process with a logistic loss $\ell(q)=\log \left(1+e^{-q}\right)$ can be viewed as an implicit margin maximization problem.

As we discussed in Theorem 1, for a homogeneous ReLU neural network $M(\boldsymbol{\theta} ; \cdot)$, gradient flow on the logistic loss 

\begin{equation}
    \mathcal{L}(\boldsymbol{\theta}):=\sum_{i=1}^n \ell\left(y_i M\left(\boldsymbol{\theta} ; \mathbf{x}_i\right)\right)
\end{equation}

converges in direction to a KKT point of:
\begin{equation}
    \min _{\boldsymbol{\theta}^{\prime}} \frac{1}{2}\left\|\boldsymbol{\theta}^{\prime}\right\|^2 
\end{equation}

subject to 

\begin{equation}
    \forall i \in[n] \quad y_i M\left(\boldsymbol{\theta}^{\prime} ; \mathbf{x}_i\right) \geq 1
\end{equation}

The KKT conditions for this problem yield~\cite{DBLP:conf/iclr/LyuL20,DBLP:conf/nips/JiT20}:

\begin{equation}
    \label{equ:original}
    \begin{array}{l}
    \tilde{\boldsymbol{\theta}}=\sum_{i=1}^n \lambda_i y_i \nabla_{\boldsymbol{\theta}} M\left(\tilde{\boldsymbol{\theta}} ; \mathbf{x}_i\right)  \\
    \forall i \in[n], y_i M\left(\tilde{\boldsymbol{\theta}} ; \mathbf{x}_i\right) \geq 1  \\
    \lambda_1, \ldots, \lambda_n \geq 0  \\
    \forall i \in[n], \quad \lambda_i=0 \text { if } y_i M\left(\tilde{\boldsymbol{\theta}} ; \mathbf{x}_i\right) \neq 1 
    \end{array}
\end{equation}

Based on these KKT conditions, we can formulate a reconstruction method~\cite{DBLP:conf/nips/HaimVYSI22,DBLP:conf/nips/BuzagloHYVONI23}. The goal is to find a set of \( \{\mathbf{x}_i\}_{i=1}^m \) and \( \{\lambda_i\}_{i=1}^m \) that satisfy following equation:

\begin{equation}
    \min_{\{\mathbf{x}_i\}_{i=1}^m, \{\lambda_i\}_{i=1}^m} L_{reconstruct} = \alpha_1 L_{stationary} + \alpha_2 L_{\lambda}
\end{equation}

where:

\begin{equation}
    \label{equ:recover_1}
    L_{stationary} = \left\| \tilde{\boldsymbol{\theta}} - \sum_{i=1}^m \lambda_i y_i \nabla_{\boldsymbol{\theta}} M\left(\tilde{\boldsymbol{\theta}}; \mathbf{x}_i\right) \right\|^2
\end{equation}
\begin{equation}
    L_{\lambda} = \sum_{i=1}^m \max\{-\lambda_i, 0\}
\end{equation}

Let $U$ be an unlearning operator that unlearns $(\mathbf{x}_i, y_i)$ from a model with parameters $\tilde{\boldsymbol{\theta}}$ to $\boldsymbol{\theta'}$:

\begin{equation}
    \boldsymbol{\theta'} = U(M, \mathcal{D}, (x_k, y_k))
\end{equation}

To verify unlearning, we compare the recovered samples before and after unlearning those samples. Let \( F \) be the index set of samples to be unlearned. The parameters of the model after unlearning \( \boldsymbol{\theta}' \)  should also satisfy the KKT conditions:

\begin{equation}
    \label{equ:after}
    \begin{array}{l}
    \boldsymbol{\theta}' = \sum_{i \notin F} \lambda_i^{'} y_i \nabla_{\boldsymbol{\theta}} M\left(\boldsymbol{\theta}'; \mathbf{x}_i\right)  \\
    \forall i \in[n], y_i M\left(\boldsymbol{\theta}' ; \mathbf{x}_i\right) \geq 1  \\
    \lambda_1^{'}, \ldots, \lambda_n^{'} \geq 0  \\
    \forall i \in[n], \quad \lambda_i^{'}=0 \text { if } y_i M\left(\boldsymbol{\theta}' ; \mathbf{x}_i\right) \neq 1 
    \end{array}
\end{equation}

And the corresponding $L_{stationary}$ is 

\begin{equation}
    \min_{\{\mathbf{x}^{'}_i\}_{i=1}^m, \{\lambda^{'}_i\}_{i=1}^m} L_{reconstruct} = \alpha_1 L_{stationary} + \alpha_2 L_{\lambda}
\end{equation}

where:

\begin{equation}
    \label{equ:recover_2}
    L_{stationary} = \left\| \boldsymbol{\theta}' - \sum_{i=1}^m \lambda_i^{'} y_i \nabla_{\boldsymbol{\theta}} M\left(\boldsymbol{\theta}'; \mathbf{x}^{'}_i\right) \right\|^2
\end{equation}
\begin{equation}
    L_{\lambda} = \sum_{i=1}^m \max\{-\lambda^{'}_i, 0\}
\end{equation}

The verification involves:
\begin{enumerate}
    \item Performing sample recovery on both the original model \( \tilde{\boldsymbol{\theta}} \) and the unlearned model \( \boldsymbol{\theta}' \), obtaining recovery samples sets \( X = \{\mathbf{x}_i\}_{i=1}^m \) and \( X' = \{\mathbf{x}'_i\}_{i=1}^m \).
    \item Calculating the differences between recovered samples:
    \begin{equation}
        D_i = \|\mathbf{x}_i - \mathbf{x}'_i\|
    \end{equation}
    \item Comparing the differences for samples that unlearned and retained samples:
    \begin{equation}
        E[D_i \mid i \in F] \text{ vs } E[D_i \mid i \notin F]
    \end{equation}
\end{enumerate}

If the unlearning process is effective, we expect:

\begin{equation}
    E[D_i \mid i \in F] > E[D_i \mid i \notin F]
\end{equation}

This expectation arises because:
\begin{enumerate}
    \item For \( i \notin F \), both \( \mathbf{x}_i \) and \( \mathbf{x}'_i \) should satisfy the constraints of equations~\ref{equ:original} and~\ref{equ:after}, leading to small differences.
    \item For \( i \in F \), \( \mathbf{x}_i \) satisfies equation~\ref{equ:original} but not~\ref{equ:after}, while \( \mathbf{x}'_i \) does not satisfy equation~\ref{equ:original} but satisfies equation~\ref{equ:after}, resulting in larger differences.
\end{enumerate}

Additionally, recent work proposed by Buzaglo et al.~\cite{DBLP:conf/nips/BuzagloHYVONI23} has extended the reconstruction method in~\cite{DBLP:conf/nips/HaimVYSI22} to multi-class classification scenarios. For a dataset $D_m=\{\left(\mathbf{x}_i, y_i\right)\}_{i=1}^n \subseteq \mathbb{R}^d \times [C]$, where $C$ is the number of classes, and a neural network $M(\boldsymbol{\theta} ; \cdot)$: $\mathbb{R}^d \rightarrow \mathbb{R}^C$.  Then, the KKT conditions for the multi-class problem yield:

\begin{equation}
\begin{array}{l}
\boldsymbol{\theta} = \sum_{i=1}^n \sum_{j \neq y_i} \lambda_{i,j} \nabla_{\boldsymbol{\theta}} \left( \Delta_{y_i,j}(\boldsymbol{\theta}) \right) \\
\Delta_{y_i,j}(\boldsymbol{\theta}) = M_{y_i}(\boldsymbol{\theta} ; \mathbf{x}_i) - M_j(\boldsymbol{\theta} ; \mathbf{x}_i)\\
\forall i \in[n], \forall j \in[C] \backslash\left\{y_i\right\}:  \Delta_{y_i,j}(\boldsymbol{\theta}) \geq 1 \\
\forall i \in[n], \forall j \in[C] \backslash\left\{y_i\right\}: \lambda_{i, j} \geq 0 \\
\forall i \in[n], \forall j \in[C] \backslash\left\{y_i\right\}:  \lambda_{i, j}=0 \text { if } \Delta_{y_i,j}(\boldsymbol{\theta}) \neq 1
\end{array}
\end{equation}

The corresponding loss $L_{stationary}$ can be formulated as:

\begin{equation}
\label{equ:multi_recover}
\left| \boldsymbol{\theta} - \sum_{i=1}^m \lambda_i \nabla_{\boldsymbol{\theta}} [M_{y_i}\left(\boldsymbol{\theta}; \mathbf{x}i\right) - \max_{j \neq y_i} M_j\left(\boldsymbol{\theta}; \mathbf{x}_i\right)] \right|^2
\end{equation}

\begin{figure*}[!t]
    \centering
    \begin{subfigure}[b]{0.24\linewidth}
        \centering
        \includegraphics[width=\textwidth]{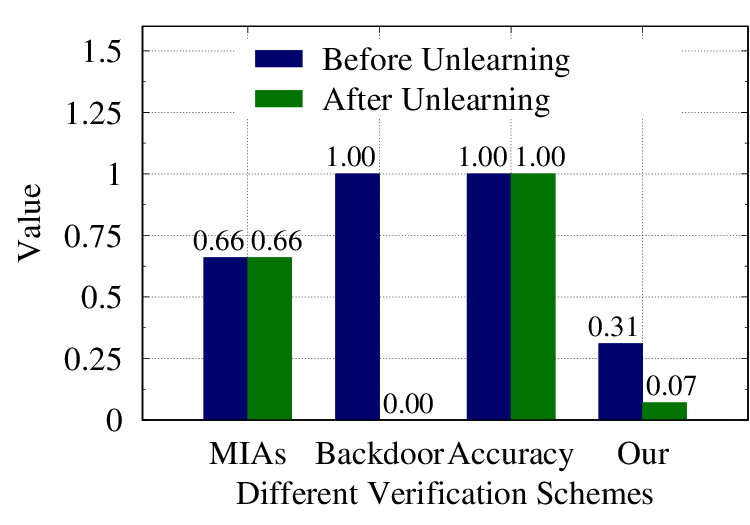}
        \caption{MNIST}
        \label{fig:sample_unlearning_mnist}
    \end{subfigure}
    \hfill
    \begin{subfigure}[b]{0.24\linewidth}
        \centering
        \includegraphics[width=\textwidth]{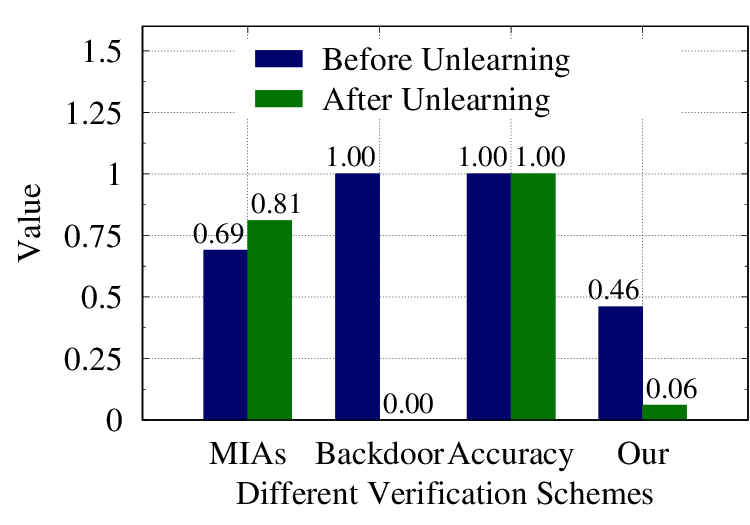}
        \caption{FMNIST}
        \label{fig:sample_unlearning_fmnist}
    \end{subfigure}
    \hfill
    \begin{subfigure}[b]{0.24\linewidth}
        \centering
        \includegraphics[width=\textwidth]{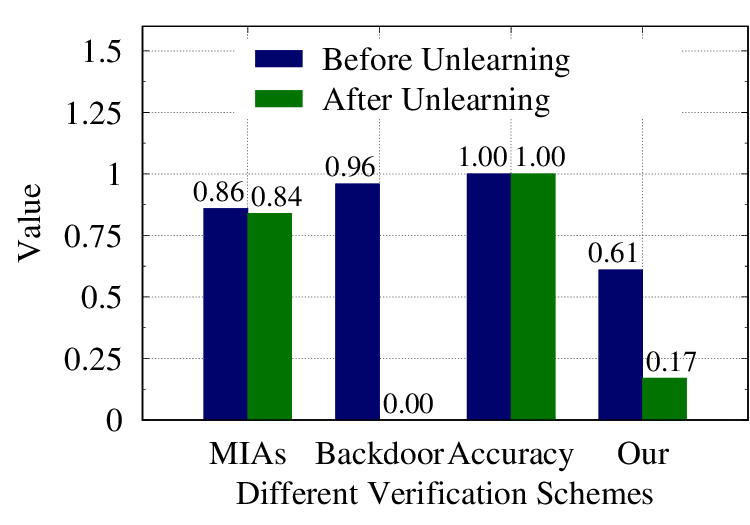}
        \caption{CIFAR-10}
        \label{fig:sample_unlearning_cifar10}
    \end{subfigure}
    \hfill
    \begin{subfigure}[b]{0.24\linewidth}
        \centering
        \includegraphics[width=\textwidth]{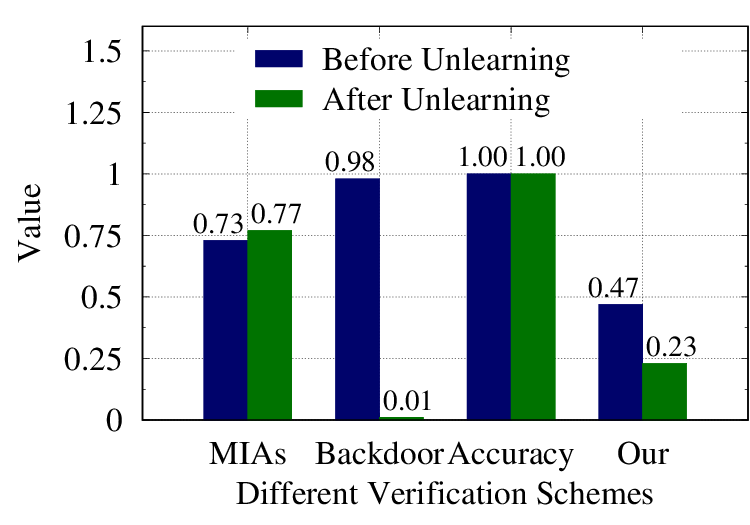}
        \caption{SVHN}
        \label{fig:sample_unlearning_svhn}
    \end{subfigure}
    \caption{Verification results for unlearning samples under sample-level unlearning requests across different schemes. The value on the Y-axis represents different metrics depending on the X-axis: INA for membership inference attacks-based scheme~(MIAs), ASR for backdoor-based schemes~(Backdoor), accuracy for accuracy-based schemes~(Accuracy), and SSIM for our proposed scheme~(Our). Only the backdoor-based scheme and our scheme show significant changes regarding the unlearning samples.}
    \label{fig:sample_unlearning_unlearning}
\end{figure*}

\begin{figure*}[!t]
    \centering
    \begin{subfigure}[b]{0.24\linewidth}
        \centering
        \includegraphics[width=\textwidth]{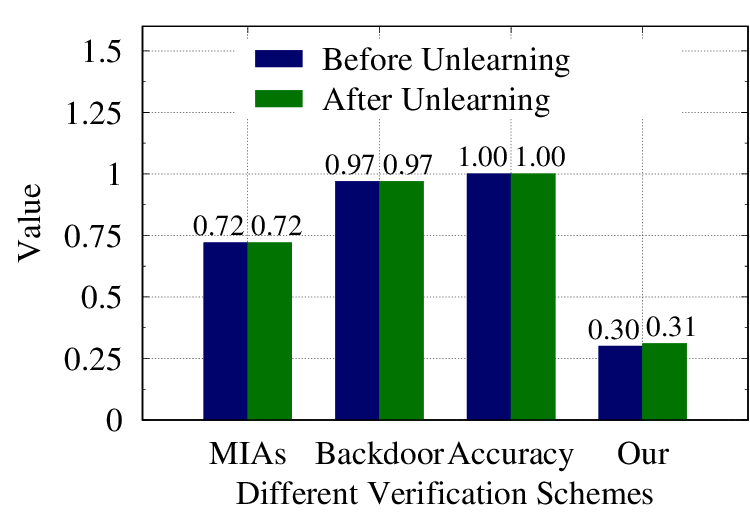}
        \caption{MNIST}
        \label{fig:sample_remaining_mnist}
    \end{subfigure}
    \hfill
    \begin{subfigure}[b]{0.24\linewidth}
        \centering
        \includegraphics[width=\textwidth]{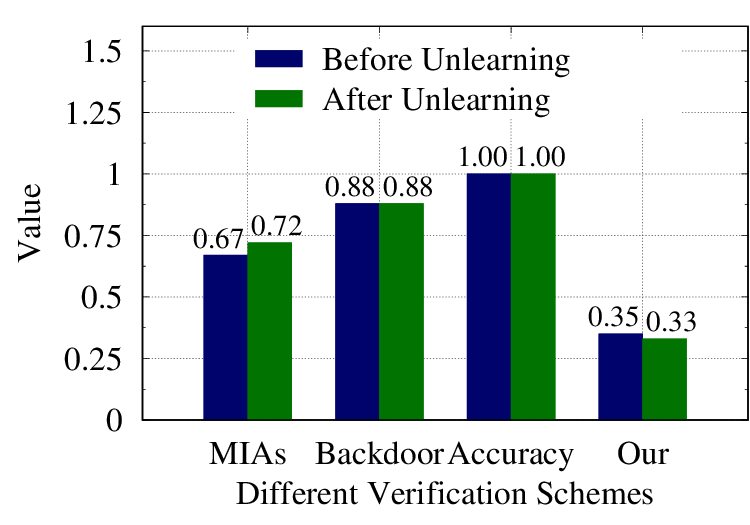}
        \caption{FMNIST}
        \label{fig:sample_remaining_fmnist}
    \end{subfigure}
    \hfill
    \begin{subfigure}[b]{0.24\linewidth}
        \centering
        \includegraphics[width=\textwidth]{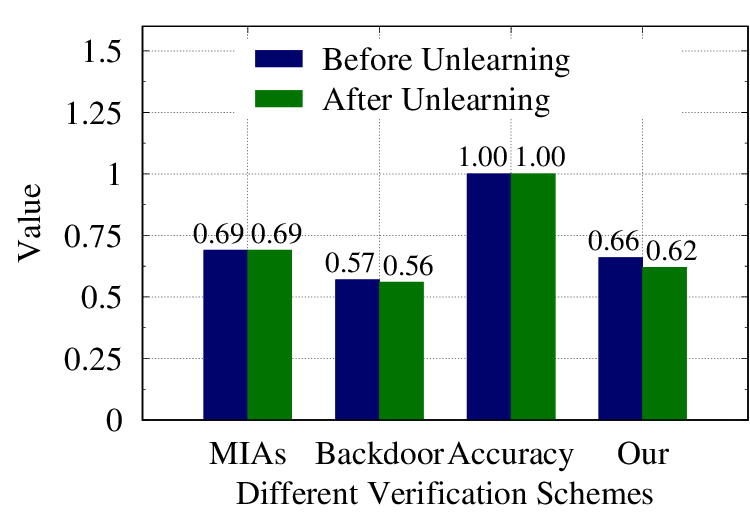}
        \caption{CIFAR-10}
        \label{fig:sample_remaining_cifar10}
    \end{subfigure}
    \hfill
    \begin{subfigure}[b]{0.24\linewidth}
        \centering
        \includegraphics[width=\textwidth]{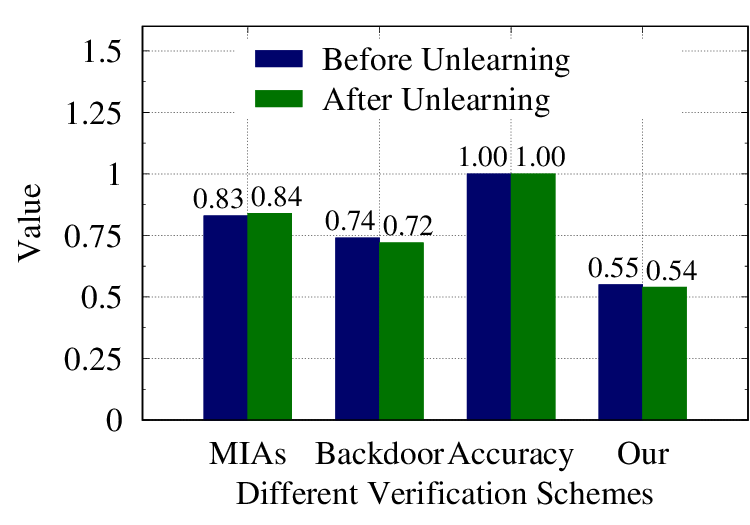}
        \caption{SVHN}
        \label{fig:sample_remaining_svhn}
    \end{subfigure}
    
    \caption{Verification results for remaining samples under sample-level unlearning requests across different schemes. Y-axis also represents different metrics depending on the X-axis: INA for MIAs-based, ASR for backdoor-based, accuracy for accuracy-based, and SSIM for our proposed scheme. All schemes do not show significant changes for the remaining samples.}
    \label{fig:sample_unlearning_remaining}
\end{figure*}

\begin{figure}[!t]
    \centering
    \begin{subfigure}[b]{0.3\linewidth}
        \centering
        \includegraphics[width=\textwidth]{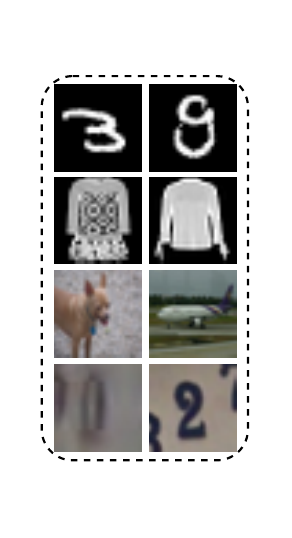}
        \caption{Original Sample}
        \label{fig:sample_orginal}
    \end{subfigure}
    \begin{subfigure}[b]{0.3\linewidth}
        \centering
        \includegraphics[width=\textwidth]{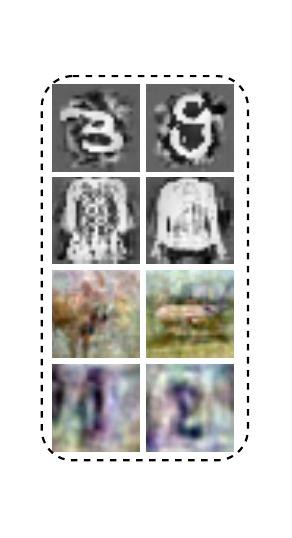}
        \caption{Before Unlearning}
        \label{fig:sample_before}
    \end{subfigure}
    \begin{subfigure}[b]{0.3\linewidth}
        \centering
        \includegraphics[width=\textwidth]{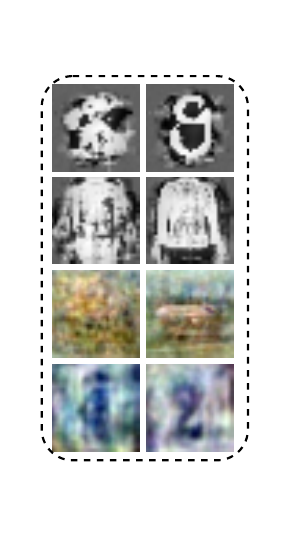}
        \caption{After Unlearning}
        \label{fig:sample_after}
    \end{subfigure}
    \caption{Original and recovered samples under the sample-level unlearning requests. Figure~\ref{fig:sample_orginal} shows two sets: original unlearning samples on the left and remaining samples on the right. Figures~\ref{fig:sample_before} and~\ref{fig:sample_after} present the recovered samples before and after the unlearning process, respectively.}
    \label{fig:recovered_samples_level}
\end{figure}

In this case, the above-proof process is also applicable. We will use this loss in our experimental evaluation when dealing with multi-classification tasks.

This theoretical framework utilizes the implicit bias inherent in neural network training and Karush-Kuhn-Tucker (KKT) conditions of margin maximization to evaluate the unlearning process. By analyzing the difference between recovered samples before and after the unlearning process, we can determine if the model provider has successfully removed the targeted samples from the model.

\section{Performance Evaluation}
\subsection{Experiment Setup}
To evaluate our scheme, we utilize four widely-used image datasets: MNIST\footnote{http://yann.lecun.com/exdb/mnist/}, Fashion MNIST\footnote{http://fashion-mnist.s3-website.eu-central-1.amazonaws.com/}, CIFAR-10\footnote{https://www.cs.toronto.edu/~kriz/cifar.html} and SVHN\footnote{http://ufldl.stanford.edu/housenumbers/}. 

\subsubsection{Baseline Methods} We compare our scheme against the following established methods:
\begin{itemize}
    \item \textbf{Membership Inference Attacks-Based Scheme}: We employ the approach proposed in~\cite{DBLP:conf/uss/LiuWH000CF022} and ~\cite{DBLP:conf/aaai/GravesNG21}.
    \item \textbf{Backdoor-Based Scheme}: We adopt the scheme proposed by Li et al.~\cite{DBLP:journals/tifs/LiZYJWX23} as a baseline method.
    \item \textbf{Model Inversion Attack-Based Scheme}: We adopt the scheme proposed by Graves et al.~\cite{DBLP:conf/aaai/GravesNG21}.
    \item \textbf{Accuracy-Based Scheme}: We also compare the accuracy of our unlearned model against the original model on both the unlearning and remaining samples~\cite{DBLP:conf/icml/BrophyL21,DBLP:conf/www/Wang0XQ22}. 
\end{itemize} 

\begin{figure*}[!t]
    \centering
    \begin{subfigure}[b]{0.24\linewidth}
        \centering
        \includegraphics[width=\textwidth]{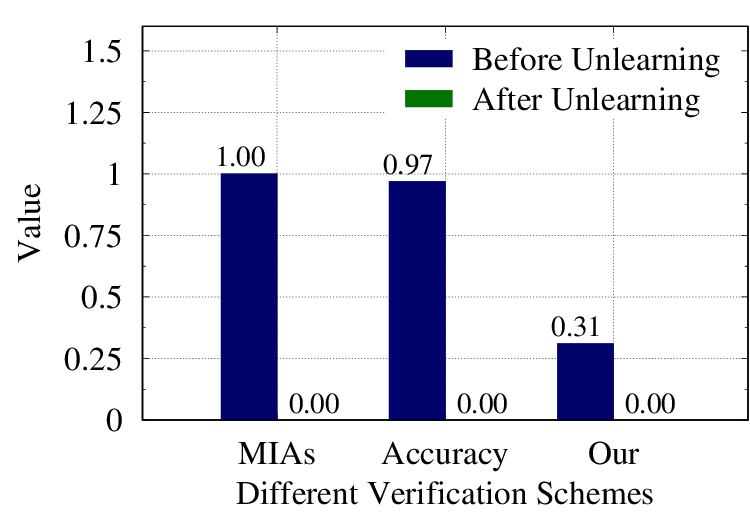}
        \caption{MNIST~(Unlearning Class)}
        \label{fig:class_unlearning_mnist}
    \end{subfigure}
    \begin{subfigure}[b]{0.24\linewidth}
        \centering
        \includegraphics[width=\textwidth]{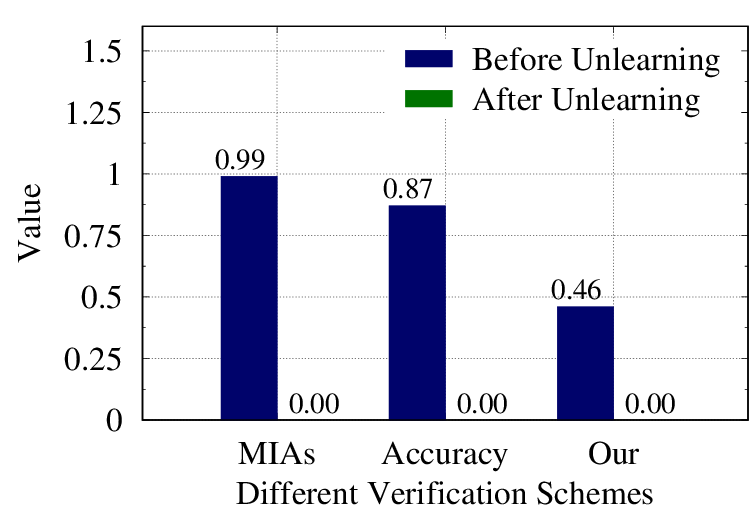}
        \caption{FashionMNIST~(Unlearning Class)}
        \label{fig:class_unlearning_fmnist}
    \end{subfigure}
    \begin{subfigure}[b]{0.24\linewidth}
        \centering
        \includegraphics[width=\textwidth]{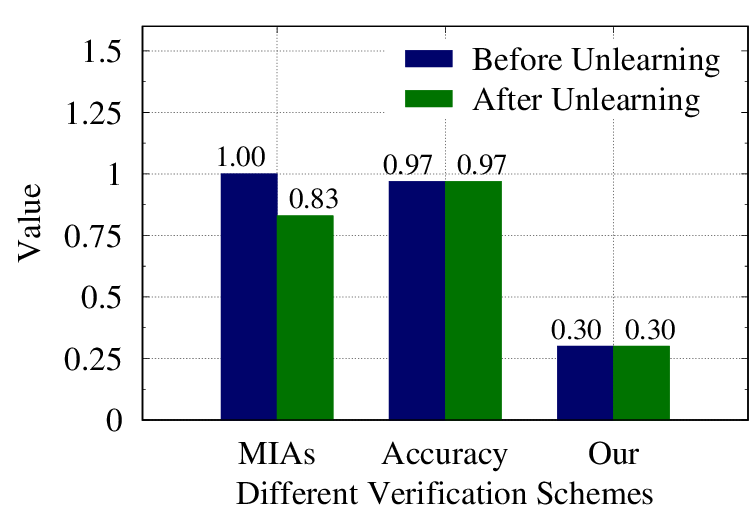}
        \caption{MNIST~(Remaining Class)}
        \label{fig:class_remaining_mnist}
    \end{subfigure}
    \begin{subfigure}[b]{0.24\linewidth}
        \centering
        \includegraphics[width=\textwidth]{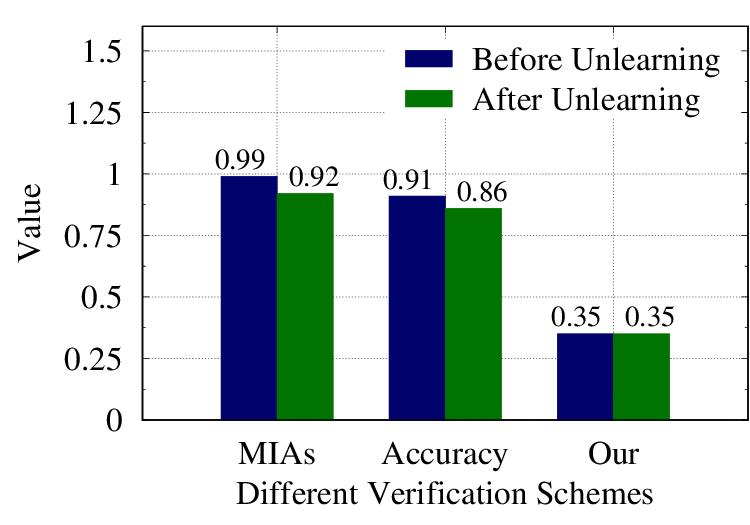}
        \caption{FashionMNIST~(Remaining Class)}
        \label{fig:class_remaining_fmnist}
    \end{subfigure}
    \caption{Verification results under class-level unlearning requests. The Y-axis in each subfigure denotes a metric specific to the X-axis variable: INA for the MIAs-based scheme, accuracy for the accuracy-based scheme, and SSIM for our scheme. All methods are effective for class-level unlearning verification.}
    \label{fig:class_unlearning}
\end{figure*}

\begin{figure}[!t]
    \centering
    \begin{subfigure}[b]{0.45\linewidth}
        \centering
        \includegraphics[width=\textwidth]{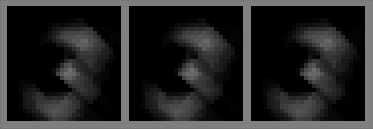}
        \caption{Before Unlearning}
        \label{fig:miv_before}
    \end{subfigure}
    \begin{subfigure}[b]{0.45\linewidth}
        \centering
        \includegraphics[width=\textwidth]{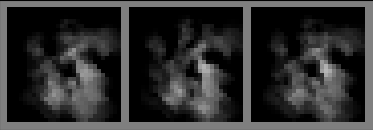}
        \caption{After Unlearning}
        \label{fig:miv_after}
    \end{subfigure}
    \caption{Reconstructed class representatives based on model inversion attack before and after unlearning in the class-level unlearning setting. Figure~\ref{fig:miv_before} shows the reconstructed class representatives before the unlearning process, while Figure~\ref{fig:miv_after} shows the corresponding samples after the unlearning process.}
    \label{fig:reconstructed_miv}
\end{figure}

\subsubsection{Metrics} We consider the following metrics to separately evaluate the baseline scheme and our scheme:
\begin{itemize}
    \item For membership inference attacks-based scheme~\cite{DBLP:conf/uss/LiuWH000CF022}, we use the success rate of attacking samples as our metric, defined as $INA = \frac{TP}{TP + FN}$ where $TP$ is the number of samples predicted to be in the training set, and $TP + FN$  is the total tested samples. Ideally, $INA$ should be close to $100\%$ before unlearning and approach $0\%$ after unlearning. 
    \item For backdoor-based scheme~\cite{DBLP:journals/tifs/LiZYJWX23}, we also use the attack success rate, $ASR$, used in~\cite{DBLP:journals/tifs/LiZYJWX23} to evaluate the unlearning results. Ideally, before unlearning, $ASR$ should be close to $100\%$. After unlearning, $ASR$ should approach $0\%$.
    \item For model inversion attack-based scheme~\cite{DBLP:conf/aaai/GravesNG21}, we directly show the recovered samples. Ideally, before unlearning, those samples should contain discernible information about the class targeted for unlearning. After unlearning, Those samples should appear dark, jumbled, and dissimilar from the unlearning class, indicating the successful unlearning of class-specific information.
    \item For our scheme, we evaluate it using both qualitative and quantitative perspectives. Qualitative: visual inspection of recovered samples, similar to the model inversion attack-based schemes. Quantitative: we use the SSIM to evaluate the similarity between recovered and original unlearning samples. SSIM values range from $-1$ to $1$, with higher values indicating greater similarity.

\end{itemize}

\begin{figure}[!t]
    \centering
    \begin{subfigure}[b]{0.3\linewidth}
        \centering
        \includegraphics[width=\textwidth]{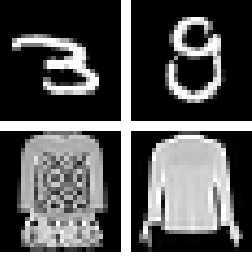}
        \caption{Original Sample}
        \label{fig:class_orginal}
    \end{subfigure}
    \begin{subfigure}[b]{0.3\linewidth}
        \centering
        \includegraphics[width=\textwidth]{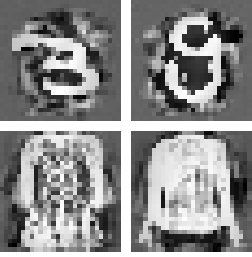}
        \caption{Before Unlearning}
        \label{fig:class_before}
    \end{subfigure}
    \begin{subfigure}[b]{0.3\linewidth}
        \centering
        \includegraphics[width=\textwidth]{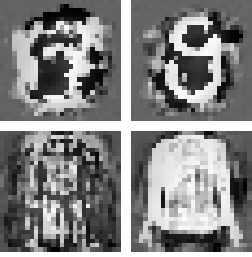}
        \caption{After Unlearning}
        \label{fig:class_after}
    \end{subfigure}
    \caption{Original and recovered samples based on our scheme under the class-level unlearning request. Figure~\ref{fig:class_orginal} shows some original samples in each class, with two samples from the unlearning class on the left and two samples of the remaining class on the right. Figure~\ref{fig:class_before} shows the recovered samples prior to unlearning, while Figure~\ref{fig:class_after} shows the most similar samples after the unlearning process.}
    \label{fig:recovered_class_level}
\end{figure}

\subsection{Verification Results}
\label{sec:verificationresults}
Our evaluation includes sample and class level unlearning requests. To ensure consistency with existing studies~\cite{DBLP:journals/tifs/GuoZHWJ24}, we employ retraining from scratch as our unlearning method. 

\subsubsection{Sample-Level Unlearning Verification}
\label{sec:sample_level_unlearning_verification}

Figure~\ref{fig:sample_unlearning_unlearning} and~\ref{fig:sample_unlearning_remaining} show our experimental results. Figure~\ref{fig:sample_unlearning_unlearning} illustrates the results for unlearning samples before and after the unlearning process, while Figure~\ref{fig:sample_unlearning_remaining} shows results for samples not unlearned. Each sub-figure demonstrates the performance of various verification schemes across different datasets. The Y-axis represents different metrics depending on the scheme: INA for membership inference attacks (MIAs)-based schemes, ASR for backdoor-based schemes, accuracy for accuracy-based schemes, and SSIM for our proposed verification scheme. The conclusion can be summary as follows:

\begin{itemize}
    \item MIAs-based and Accuracy-based Schemes: Before unlearning, success rates for identifying both unlearning and remaining samples are high, indicating that MIAs effectively identify samples used in model training. However, after unlearning, there's no significant reduction in attack performance for either sample type. This suggests that MIAs cannot reliably distinguish whether training samples have been successfully unlearned. Similar results can be observed from the results of accuracy-based schemes.
    \item Backdoor-based Scheme: Following backdoor embedding, the accuracy of classifying both unlearning and remaining samples as targets is very high. After unlearning, the classification accuracy for unlearning samples with the backdoor drops to nearly $0\%$, while remaining higher for other samples. This indicates that the backdoor-based verification scheme effectively confirms the unlearning process, demonstrating the successful removal of backdoor samples associated with unlearning.
    \item Our Proposed Scheme: Prior to unlearning, all recovered samples show high similarity based on the SSIM metric. After the unlearning process, SSIM values for unlearning samples decrease significantly, while remaining high for other samples. Those SSIM values before and after unlearning demonstrate that our scheme effectively distinguishes between unlearned and retained samples.
\end{itemize}

\begin{figure*}[!t]
    \centering
    \begin{subfigure}[b]{0.3\linewidth}
        \centering
        \includegraphics[width=\textwidth]{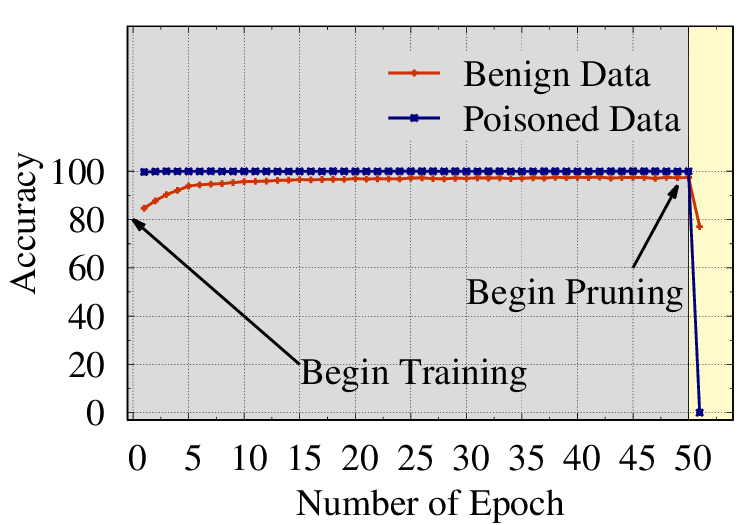}
        \caption{Results of Setting (1)}
        \label{fig:backdoor_pruning}
    \end{subfigure}
    \begin{subfigure}[b]{0.3\linewidth}
        \centering
        \includegraphics[width=\textwidth]{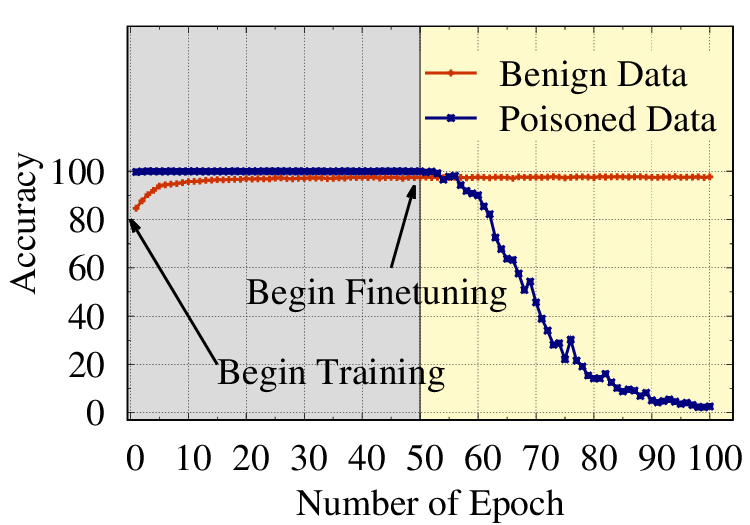}
        \caption{Results of Setting (2)}
        \label{fig:backdoor_finetuning}
    \end{subfigure}
    \begin{subfigure}[b]{0.3\linewidth}
        \centering
        \includegraphics[width=\textwidth]{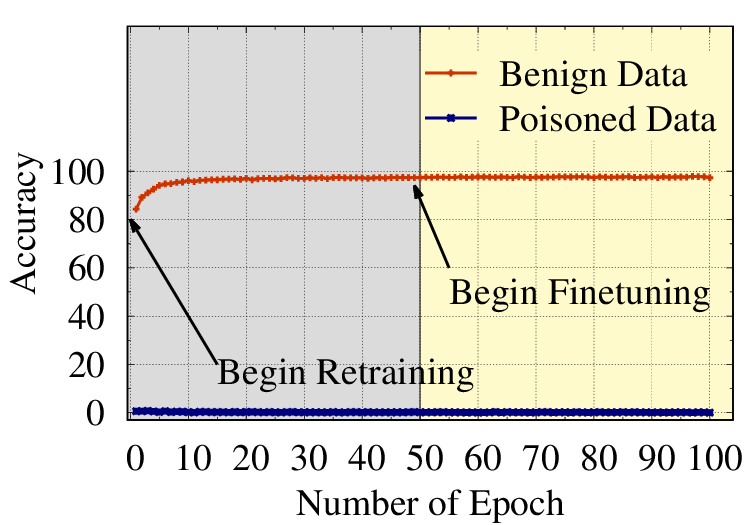}
        \caption{Results of Setting (3)}
        \label{fig:backdoor_unlearning_finetuning}
    \end{subfigure}
    \caption{Evaluation results of robustness for backdoor-based verification scheme. Pruning, directly fine-tuning, and fine-tuning after unlearning all compromise backdoor verification's effectiveness.}
    \label{fig:robustness_backdoor}
\end{figure*}

We also show some original training samples and recovered samples extracted using our scheme in Figure~\ref{fig:recovered_samples_level}. Each column shows unlearning samples on the left and the remaining samples on the right. Before unlearning, our scheme effectively recovers all samples. After unlearning, the samples that were most similar to those samples targeted for unlearning no longer contained relevant information, while for the remaining samples, our scheme still recovers images similar to the originals. This demonstrates our scheme's ability to effectively distinguish between unlearned and retained samples, validating it can be used as a verification method.

\subsubsection{Class-Level Unlearning Verification}
\label{sec:class_level_unlearning_verification}


We set the remaining class to $9$ for the MNIST dataset and to $0$ for the FashionMNIST dataset. The unlearning class is selected as $3$ for both datasets. Figure~\ref{fig:class_unlearning} shows the quantitative results of our evaluation. Specifically, Figure~\ref{fig:class_unlearning_mnist} and Figure~\ref{fig:class_remaining_mnist} show the results for the unlearning class and the remaining class in the MNIST dataset, while  Figure~\ref{fig:class_unlearning_fmnist} and Figure ~\ref{fig:class_remaining_fmnist} illustrate the changes in the unlearning class and the remaining class for the FashionMNIST dataset.

\begin{figure}[!t]
    \centering
    \begin{subfigure}[b]{0.19\linewidth}
        \centering
        \includegraphics[width=0.9\textwidth]{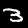}
        \caption{Original}
        \label{fig:robustness_original}
    \end{subfigure}
    \begin{subfigure}[b]{0.19\linewidth}
        \centering
        \includegraphics[width=0.9\textwidth]{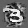}
        \caption{Baseline}
        \label{fig:robustness_baseline}
    \end{subfigure}
    \begin{subfigure}[b]{0.19\linewidth}
        \centering
        \includegraphics[width=0.9\textwidth]{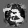}
        \caption{Setting (1)}
        \label{fig:robustness_pruning}
    \end{subfigure}
    \begin{subfigure}[b]{0.19\linewidth}
        \centering
        \includegraphics[width=0.9\textwidth]{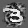}
        \caption{Setting (2)}
        \label{fig:robustness_finetuning}
    \end{subfigure}
    \begin{subfigure}[b]{0.19\linewidth}
        \centering
        \includegraphics[width=0.9\textwidth]{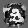}
        \caption{Setting (3)}
        \label{fig:robutsness_unlearning_finetuning}
    \end{subfigure}
    \caption{Evaluation results of robustness for our scheme. Our scheme effectively recovers the training samples even after the model has been subjected to pruning, fine-tuning, and unlearning followed by fine-tuning. This indicates that our method exhibits a certain degree of robustness.}
    \label{fig:robustness_our}
\end{figure}

\textbf{Results.} For samples in the class that need to be unlearned, MIA-based, accuracy-based, and our proposed methods all show significant changes~(see Figures~\ref{fig:class_unlearning_mnist} and~\ref{fig:class_unlearning_fmnist}). For remaining class, no significant changes are observed across all three approaches~(see Figures~\ref{fig:class_remaining_mnist} and~\ref{fig:class_remaining_fmnist}). This indicates that all methods are effective for class-level unlearning verification.

Figure~\ref{fig:reconstructed_miv} and~\ref{fig:recovered_class_level} show some original and recovered samples extracted using the model inversion attack-based scheme and our scheme. In Figure~\ref{fig:reconstructed_miv}, before unlearning, the model inversion attack reconstructs representatives containing information about the unlearning class. After unlearning, it produces dark, jumbled images with almost no information about the unlearning class. Similarly, for our scheme, the recovered sample after unlearning all samples in the unlearning class also becomes dark and jumbled, which illustrates that our scheme can also be used in verifying class-level unlearning requests.

\textbf{Summary.} The above experiments demonstrate that the proposed verification scheme effectively validates both sample-level and class-level unlearning requests. The method consistently performs well across these different granularity levels, showcasing its applicability and reliability in various scenarios.

\subsection{Robustness Evaluation}

As highlighted in Section~\ref{sec:existenceschemes}, the robustness of unlearning verification scheme is crucial for long-term deployment. A robust verification scheme should remain effective even after fine-tuning or pruning the models when the training or unlearning process is finished. In this Section, we evaluate the robustness of our proposed scheme, comparing it with the backdoor-based verification scheme~\cite{DBLP:journals/tifs/LiZYJWX23}. We use the same experimental settings described in Section~\ref{sec:sample_level_unlearning_verification} and re-evaluate these schemes after performing the following operations:

\begin{enumerate}[label=(\arabic*)]
    \item Training model $\rightarrow$  Pruning model
    \item Training model $\rightarrow$  Fine-tuning model
    \item Training model $\rightarrow$  Unlearning $\rightarrow$  Fine-tuning model
\end{enumerate}

For pruning, we randomly prune $20\%$ of the parameters in each layer. For fine-tuning, we train the model using the original training samples and correct labels, maintaining the initial training configuration. For unlearning, we use retraining from scratch ~\cite{DBLP:journals/tifs/GuoZHWJ24}. The results of the backdoor-based scheme are illustrated in Figure~\ref{fig:robustness_backdoor}, while Figure~\ref{fig:robustness_our} presents the results of our proposed scheme.

\begin{figure*}[!t]
    \centering
    \begin{subfigure}[b]{0.3\linewidth}
        \centering
        \includegraphics[width=\textwidth]{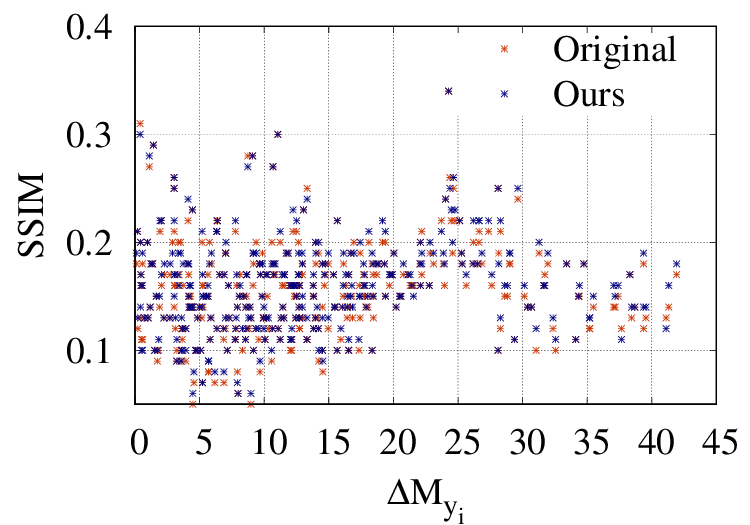}
        \caption{SSIM Distribution}
        \label{fig:partial_samples}
    \end{subfigure}
    \begin{subfigure}[b]{0.3\linewidth}
        \centering
        \includegraphics[width=\textwidth]{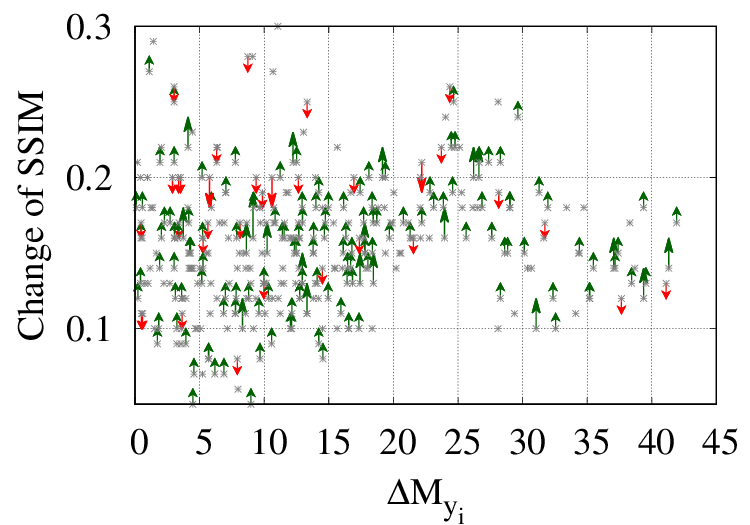}
        \caption{SSIM Change}
        \label{fig:partialarrow}
    \end{subfigure}
    \begin{subfigure}[b]{0.3\linewidth}
        \centering
        \includegraphics[width=\textwidth]{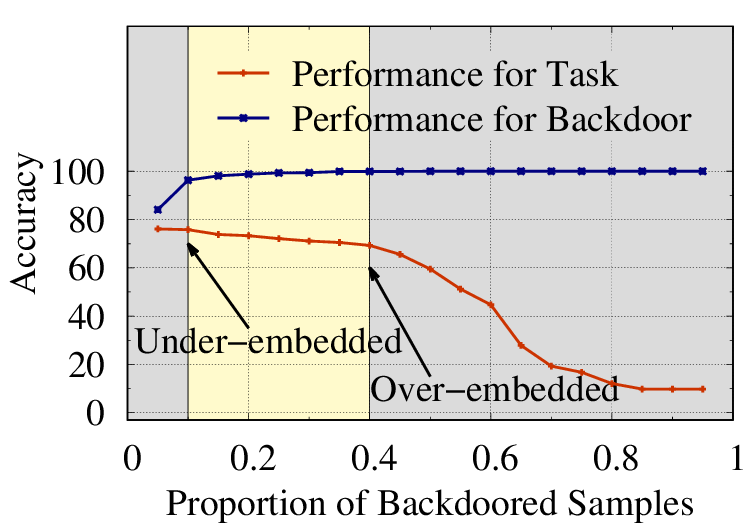}
        \caption{Backdoor-based Scheme}
        \label{fig:partial_backdoor}
    \end{subfigure}
    \caption{Evaluation results of the recovery quality (measured by SSIM) against the sample's distance from the decision boundary. Comparing the scheme in~\cite{DBLP:conf/nips/BuzagloHYVONI23}, our scheme shows an improvement, with a larger proportion of recovered samples exhibiting greater similarity to the original training samples. Figure~\ref{fig:partialarrow} visually illustrated this phenomenon. In addition, our scheme can support over 40 times the number of verifications compared to the backdoor-based method~\cite{DBLP:journals/tifs/LiZYJWX23}.}
    \label{fig:ablation_studyand_partial_analysis}
\end{figure*}

\textbf{Results}: As shown in Figure~\ref{fig:backdoor_pruning}, when the model undergoes pruning after training, the backdoor-based verification scheme becomes ineffective, as evidenced by the near-zero performance on the poison dataset. This indicates that the backdoor-based verification loses its robustness in the face of pruning operations, rendering it unusable. Similarly, in Figure~\ref{fig:backdoor_finetuning}, when fine-tuning is performed using the original, unperturbed training samples, the accuracy of the backdoor also decreases, indicating that fine-tuning also compromises the robustness of the backdoor-based verification scheme. Lastly, Figure~\ref{fig:backdoor_unlearning_finetuning} illustrates that when the model is finetuned with previously unlearned samples after unlearning, the backdoor-based verification scheme also fails due to the absence of the backdoor pattern in those samples. In conclusion, the backdoor-based method relies on pre-prepared samples for verification. If the backdoor pattern is not embedded in advance, if those patterns are disrupted after training, or if the backdoor has been previously used, subsequent verification will be infeasible.

Figure~\ref{fig:robustness_our} shows the results obtained from our scheme. Figure~\ref{fig:robustness_original} illustrate the original training sample, while other Figures show samples recovered after various processes: initial training (Figure~\ref{fig:robustness_baseline}), pruning (Figure~\ref{fig:robustness_pruning}), fine-tuning (Figure~\ref{fig:robustness_finetuning}), and unlearning followed by fine-tuning (Figure~\ref{fig:robutsness_unlearning_finetuning}). Unlike the backdoor-based scheme, our scheme effectively recovers training samples even after the model has been subjected to these modifications. This consistent performance demonstrates that our method exhibits a higher degree of robustness, maintaining its effectiveness across various post-training adjustments where the backdoor-based scheme fails.

\begin{figure}[!t]
    \centering
    \begin{subfigure}[b]{0.3\linewidth}
        \centering
        \includegraphics[width=\textwidth]{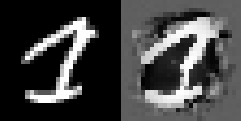}
        \caption{SSIM=0.3}
    \end{subfigure}
    \begin{subfigure}[b]{0.3\linewidth}
        \centering
        \includegraphics[width=\textwidth]{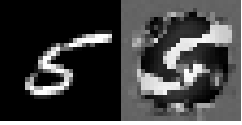}
        \caption{SSIM=0.2}
    \end{subfigure}
    \begin{subfigure}[b]{0.3\linewidth}
        \centering
        \includegraphics[width=\textwidth]{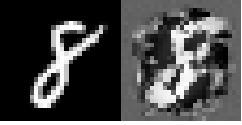}
        \caption{SSIM=0.15}
    \end{subfigure}
    \caption{Evaluation results of verification range. Recovered samples were selected based on various values of SSIMs.}
    \label{fig:ssim_partial_analysis}
\end{figure}

\textbf{Summary.} The above experiments demonstrate that, as the backdoor-based verification scheme depends on pre-prepared patterns for verification, if those pre-prepared patterns are not embedded in advance, are disrupted after training, or have been previously used, subsequent verification becomes infeasible. In contrast, our method demonstrates consistent performance and a higher degree of robustness, maintaining effectiveness across various post-training adjustments.

\subsection{Ablation Study and Analysis of Verification Range}
As discussed in Section~\ref{sec:existenceschemes}, the ability to verify more samples is crucial in MLaaS unlearning verification. This section evaluates the number of verifications supported by our proposed scheme compared to the method introduced in~\cite{DBLP:journals/tifs/LiZYJWX23}. Furthermore, in Section~\ref{sec:datareconstruction}, we add a new constraint to improve the quality of recovered samples and provide more samples used for unlearning verification. In this Section, we also do a comparative analysis between our enhanced scheme and the data reconstruction scheme proposed in~\cite{DBLP:conf/nips/BuzagloHYVONI23}.

We follow experimental settings used in~\cite{DBLP:conf/nips/BuzagloHYVONI23} to construct our comparison experiment. Specifically, we use the experimental setting in ~\ref{sec:sample_level_unlearning_verification} and select the MNIST dataset. We first recover samples based on the loss $L_{\text {reconstruct }} =  \alpha_1 L_{\text {stationary }}+\alpha_2 L_\lambda$ and record the result as original. Then we add our new loss $\alpha_3 L_{prior}$ to $L_{\text {reconstruct }}$ and continue to recover samples. We set $\alpha_1 = 1$, $\alpha_1 = 1$ and $\alpha_3 =1$ and record the recover samples as ours. We adopted the same other hyperparameters provided in~\cite{DBLP:conf/nips/BuzagloHYVONI23}. To evaluate the quality of our recovered samples, we employ the same evaluation method in~\cite{DBLP:conf/nips/BuzagloHYVONI23}: for each sample in the original training dataset we search for its nearest neighbor in the recovered samples and measure the similarity using SSIM. A higher SSIM value indicates better recovery quality. In Figure~\ref{fig:partial_samples}, we plot the recovery quality (measured by SSIM) against the sample's distance from the decision boundary. This distance is calculated based on: 

$$\Delta M_{y_i} = M_{y_i}(\mathbf{x}_i; \boldsymbol{\theta}) - \max_{j \neq y_i} M_j(\mathbf{x}_i; \boldsymbol{\theta})$$

where $M_{y_i}(\mathbf{x}_i; \boldsymbol{\theta})$ is the logit for the true class, and
$\max_{j \neq y_i} M_j(\mathbf{x}_i; \boldsymbol{\theta})$ is the maximum logit among all other classes. In Figure~\ref{fig:partialarrow}, we show the change of SSIM for each corresponding original sample. For the backdoor-based scheme~\cite{DBLP:journals/tifs/LiZYJWX23}, we choose the experimental setting in~\ref{sec:sample_level_unlearning_verification} and use the same number of the training dataset in Buzaglo et al.~\cite{DBLP:conf/nips/BuzagloHYVONI23}. The corresponding results are shown in Figure~\ref{fig:partial_backdoor}.

\begin{figure*}[!t]
    \centering
    \begin{subfigure}[t]{0.36\linewidth}
        \centering
        \includegraphics[width=\textwidth]{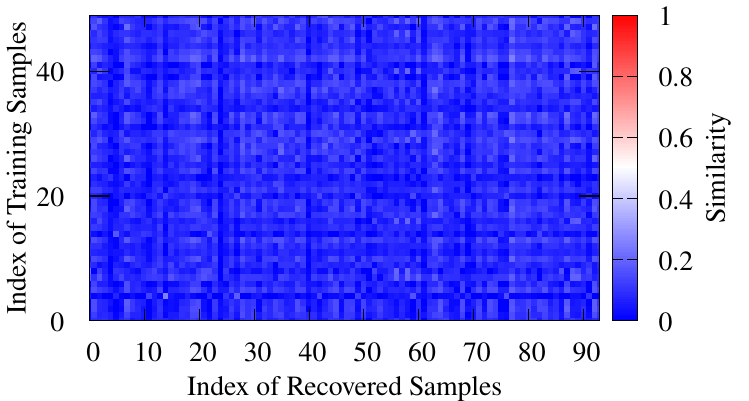}
        \caption{Before Fine-tuning}
        \label{fig:finetuning_is_not_a_original}
    \end{subfigure}
    \begin{subfigure}[t]{0.62\linewidth}
        \centering
        \includegraphics[width=\textwidth]{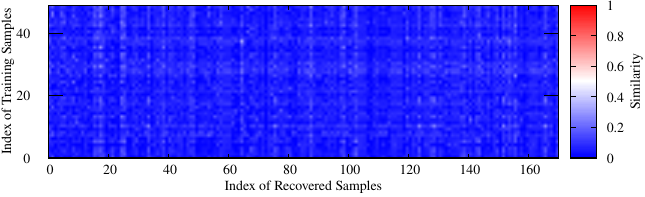}
        \caption{After Fine-tuning}
        \label{fig:finetuning_is_not_a_finetuned}
    \end{subfigure}
    \caption{Evaluation results of unlearning scheme based on relabel-based fine-tuning. Before relabel-based fine-tuning, SSIM between those recovered and training samples is very high~(see Figure~\ref{fig:finetuning_is_not_a_original}), indicating the model retains information about the class to be unlearned. After unlearning~(Figure~\ref{fig:finetuning_is_not_a_finetuned}), we can recover more samples highly similar to training samples, which concludes that the fine-tuning schemes based on random labels are not an effective unlearning solution. }
    \label{fig:finetuning_is_not_a_solution_ssim}
\end{figure*}

\textbf{Results.} Figures~\ref{fig:partial_samples} demonstrates that both schemes proposed in~\cite{DBLP:conf/nips/BuzagloHYVONI23} and ours successfully recover various samples from the model, aligning with the findings reported in~\cite{DBLP:conf/nips/BuzagloHYVONI23}. In addition, our scheme shows an improvement, with a larger proportion of recovered samples~(blue points) exhibiting greater similarity to the training dataset compared to the original scheme~(red points) in Figure~\ref{fig:partial_samples}. 
Figure~\ref{fig:partialarrow} visually represents this improvement, with green arrows indicating samples where our scheme achieves higher similarity, gray points representing unchanged similarity, and red arrows denoting decreased similarity. This improved similarity between recovered samples and training samples provides more samples for our subsequent verification processes.

\begin{figure}[!t]
    \centering
    \begin{subfigure}[t]{0.9\linewidth}
        \centering
        \includegraphics[width=\textwidth]{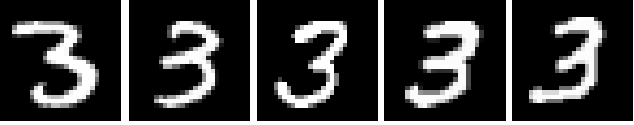}
        \caption{Original Training Samples}
        \label{fig:finetuning_is_not_a_solution}
    \end{subfigure}\\
    \begin{subfigure}[t]{0.9\linewidth}
        \centering
        \includegraphics[width=\textwidth]{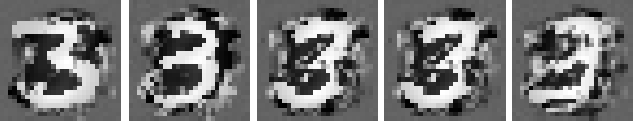}
        \caption{Recovered after Relabel-based Unlearning}
        \label{fig:finetuning_is_not_a_solution_recover}
    \end{subfigure}
    \caption{Original samples and corresponding recovered samples after random label-based funetuning. Original samples are from the training dataset, while the corresponding recovered samples are selected from those with the largest SSIM between the corresponding original samples.}
    \label{fig:finetuning_is_not_a_solution_recovered_samples}
\end{figure}

Additionally, as shown in Figure~\ref{fig:partial_backdoor}, when the proportion of backdoored samples used for training is less than $5\%$, their performance is inadequate for unlearning verification. Effective verification is achieved only when the proportion of backdoored samples reaches $10\%$. This suggests that, for the MNIST dataset, approximately $10\%$ of the total training samples (500 in total) is necessary for a single verification. Given the total number of samples, it only supports $10$ times for verification. Furthermore, as the number of backdoored samples increases, the model’s performance will significantly decrease since the samples that have not been backdoored become less. For example, from Figure~\ref{fig:partial_backdoor}, when only $60\%$ training samples for original task training, the performance of the trained model will decrease to only $44\%$. For our scheme, we show some recovered samples in Figure~\ref{fig:ssim_partial_analysis}. It can be seen that when the SSIM of the recovered image equals $0.15$, the recovered samples still partially retain information about the original sample. From Figure~\ref{fig:partial_samples}, we observe that approximately $255$ samples have an SSIM greater than $0.15$. This suggests that our scheme can support over $40$ times the number of verifications compared to backdoor-based methods, without compromising the model's original performance.

\textbf{Summary.} Our enhanced sample recovery method significantly improves the quality of samples available for verification. In addition, it also allows a substantially higher number of verifications compared to existing backdoor-based methods, while maintaining model's performance on its primary task.

\subsection{Fine-tuning is not a Solution}
\label{sec:finetuningisnotasolution}
Currently, many machine unlearning schemes achieve their goals through fine-tuning process~\cite{DBLP:journals/csur/XuZZZY24}. This typically involves manipulating the samples targeted for unlearning, such as assigning random labels, and then fine-tuning the model with those relabeled samples~\cite{DBLP:journals/hengxutbd,DBLP:journals/tnnlstarun,DBLP:conf/cvpr/ChenGL0W23,ijcai2024p40}. Experimental evaluations in these works, using membership inference attacks, model inversion attacks, backdoor attack and accuracy-based metrics, have suggested successful unlearning. However, the question remains: Is this truly the case?

Intuitively, any fine-tuning-based unlearning scheme involving unlearning samples should be considered incomplete, as samples targeted for unlearning are still processed by the model during the fine-tuning stage. To verify this hypothesis, we used the experimental setting described in Section~\ref{sec:class_level_unlearning_verification}, focusing on the MNIST dataset and replacing the unlearning method with relabel-based fine-tuning~\cite{DBLP:conf/cvpr/ChenGL0W23,ijcai2024p40}. We use our proposed scheme to recover samples both before and after fine-tuning. For evaluation, we select all recovered samples that can be correctly classified as class targeted for unlearning using the model before unlearning. Then, we calculate the SSIM between each recovered sample and its nearest original sample. Figure~\ref{fig:finetuning_is_not_a_solution_ssim} shows the SSIM results, while Figure~\ref{fig:finetuning_is_not_a_solution_recovered_samples} shows some recovered samples after fine-tuning.

\textbf{Results.} As shown in Figure~\ref{fig:finetuning_is_not_a_original}, before performing relabel-based finetuning, the SSIM between some recovered samples and training samples is very high, indicating that the trained model indeed contains some information about the class that needs to be unlearned. Figure~\ref{fig:finetuning_is_not_a_finetuned} also reveals similar results, with many recovered samples exhibiting high similarity to the training samples~(see both recovered samples in Figure~\ref{fig:finetuning_is_not_a_solution_recovered_samples}). Furthermore, after fine-tuning, the number of recovered samples unexpectedly increases, as the model retains memory not only of the original training samples but also of the newly added fine-tuning samples. All those results suggest that even after unlearning, the model still retains information about the samples that need to be unlearned. Therefore, relabel-based fine-tuning is not an effective unlearning solution.

\section{Conclusion}
In this paper, we introduce a novel approach to machine unlearning verification, addressing the challenges of prior sample-level modifications while considering both robustness and supporting verification on a much larger set. Inspired by the existing works in implicit bias and date reconstruction, we propose an optimization-based method for recovering actual training samples from models. This enables verification of unlearning by comparing samples recovered before and after the unlearning process. We provide theoretical analyses of our scheme's effectiveness. Experimental results demonstrate robust verification capabilities while supporting verify a large number of samples, marking a significant advancement in machine unlearning research. In addition, our machine unlearning verification scheme revealed that relabeling fine-tuning methods do not fully remove, but rather amplify, the influence of targeted samples, challenging previous findings. This suggests that our verification scheme can further enhance the reliability of machine unlearning.

\bibliographystyle{IEEEtran}
\bibliography{bare_jrnl_new_sample4}
\end{document}